\documentclass[a4paper,fleqn]{cas-dc}
\usepackage{amsmath,amsfonts}
\usepackage{algorithmic}
\usepackage{array}
\usepackage[caption=false,font=normalsize,labelfont=sf,textfont=sf]{subfig}
\usepackage{textcomp}
\usepackage{stfloats}
\usepackage{url}
\usepackage{verbatim}
\usepackage{graphicx}
\usepackage{xcolor}
\usepackage{tabularx} 
\usepackage{booktabs}
\usepackage[most]{tcolorbox}
\usepackage{multirow} 
\usepackage{balance}
\usepackage{listings}
\usepackage{pgfplots}
\usepackage[switch]{lineno} 
\usepackage[commandnameprefix=always]{changes}
\pgfplotsset{compat=1.18}
\usepackage{ragged2e} 
\usepackage{makecell} 
\usepackage[numbers]{natbib}

\newcolumntype{L}{>{\RaggedRight\arraybackslash}X}
\newcolumntype{P}[1]{>{\RaggedRight\arraybackslash}p{#1}}
\newcommand{\toolname}[0]{\textbf{\textsc{DeRed}}}

\def\tsc#1{\csdef{#1}{\textsc{\lowercase{#1}}\xspace}}
\tsc{WGM}
\tsc{QE}

\begin{document}

\let\WriteBookmarks\relax
\def\floatpagepagefraction{1}
\def\textpagefraction{.001}

\shorttitle{}    

\shortauthors{}  

\title [mode = title]{\toolname{}: Enhancing Third-Party Library Detection in Binaries via Deceptive Reuse Mitigation}  

\tnotemark[1] 

\tnotetext[1]{} 

%

\author[1]{Shengjia Chang}\fnmark[1]
\ead{shengjiac@bupt.edu.cn}

\author[2]{Shouguo Yang}\fnmark[2]\cormark[1]
\ead{yangsg@zgclab.edu.cn}

\author[3]{Xiao Deng}\fnmark[2]
\ead{dengx@zgclab.edu.cn}

\author[4]{Qi Meng}\fnmark[2]
\ead{mengq@zgclab.edu.cn}

\author[5]{Dan Wang}\fnmark[2]
\ead{wangdan@zgclab.edu.cn}

\author[6]{Bin Wang}\fnmark[2]
\ead{wangbin@zgclab.edu.cn}

\author[7]{Baojiang Cui}\fnmark[1]
\ead{cuibj@bupt.edu.cn}

\author[8]{Shaocong Feng}\fnmark[1]
\ead{fengshaocong@bupt.edu.cn}

\author[9]{Haoran Yu}\fnmark[1]
\ead{haoranyu@bupt.edu.cn}

\affiliation[1]{organization={Beijing University of Posts and Telecommunications},
addressline={No. 10 Xitucheng Road, Haidian District, Beijing, China}, 
city={Beijing},
postcode={100876},
country={China}}

\affiliation[2]{organization={Zhongguancun Lab},
addressline={No. 2 Cuihu Beihuan Road, Haidian District, Beijing, China}, 
city={Beijing},
postcode={100876},
country={China}}

\cortext[1]{Corresponding author}

\fntext[1]{}


\begin{abstract}
Third-Party Libraries (TPLs) are widely adopted to accelerate software development by offering reusable components. However, TPL reuse also introduces significant security risks, particularly due to the propagation of vulnerabilities. To address this, TPL detection techniques have been proposed, primarily based on syntactic and semantic feature matching. Despite their effectiveness, these methods overlook the directionality of code reuse, often leading to false positive vulnerability reports.

To overcome this limitation, this paper introduces the concept of deceptive reuse—a misleading reuse relationship between recipient software that share common TPL code—and proposes a novel approach, \toolname{}, to enhance TPL detection by distinguishing between real reuse and deceptive reuse.
\toolname{} exploits key differences between donor code areas (TPL) and recipient code areas (software) in terms of function call graph centrality and connectivity to exported functions. It further introduces algorithms to quantify these differences for accurate reuse classification. Additionally, to achieve efficient and scalable homologous code matching, \toolname{} integrates a Markov matrix and MobileNet for function matching, combined with Graph Convolutional Networks operating on Attributed Function Call Graphs for area alignment.
Experimental results demonstrate that \toolname{} effectively distinguishes between real and deceptive reuse, significantly outperforming existing TPL detection methods. Furthermore, its core components exhibit strong robustness across different architectures and compiler optimization levels. A real-world case study comparing \toolname{} with the commercial Software Composition Analysis tool, BinaryAI, demonstrates that \toolname{} outperforms BinaryAI in mitigating deceptive reuse, thereby significantly reducing false positives during vulnerability detection. The source code is available at https://github.com/kitsch0x97/dered-artifacts.
\end{abstract}


\maketitle


\section{Introduction}

With the rapid expansion of the open-source ecosystem, Third-Party Library (TPLs) have become indispensable components of modern software engineering. 
Their extensive reuse accelerates development, supported by mature integration infrastructures such as GitHub ~\cite{Github}, SourceForge ~\cite{SourceForge}, and package managers like Conan ~\cite{Conan} and VCPKG ~\cite{VCPKG}. 
According to Synopsys ~\cite{Synopsys}, 97\% of software projects incorporate at least one TPL, underscoring their central role in modern development. 
However, this widespread reuse also introduces considerable security challenges. 
Empirical studies have shown that ubiquitous libraries such as BusyBox and OpenSSL are embedded in thousands of softwares. 
For instance, BusyBox was identified in over 40\% of analyzed software samples, with the vulnerability CVE-2018-1000517 compromising thousands of devices~\cite{AutoFirm}. 
Similarly, OpenSSL appeared in more than 1,300 software from vendors such as TP-Link and D-Link, where the Heartbleed vulnerability (CVE-2014-0160) was detected in 380 instances~\cite{Empirical}. 
These findings indicate that pervasive TPL reuse can expose systems to severe vulnerabilities, highlighting the urgent need for effective TPL detection to enhance software security.
\begin{figure}
	\centering
    \includegraphics[width=\columnwidth]{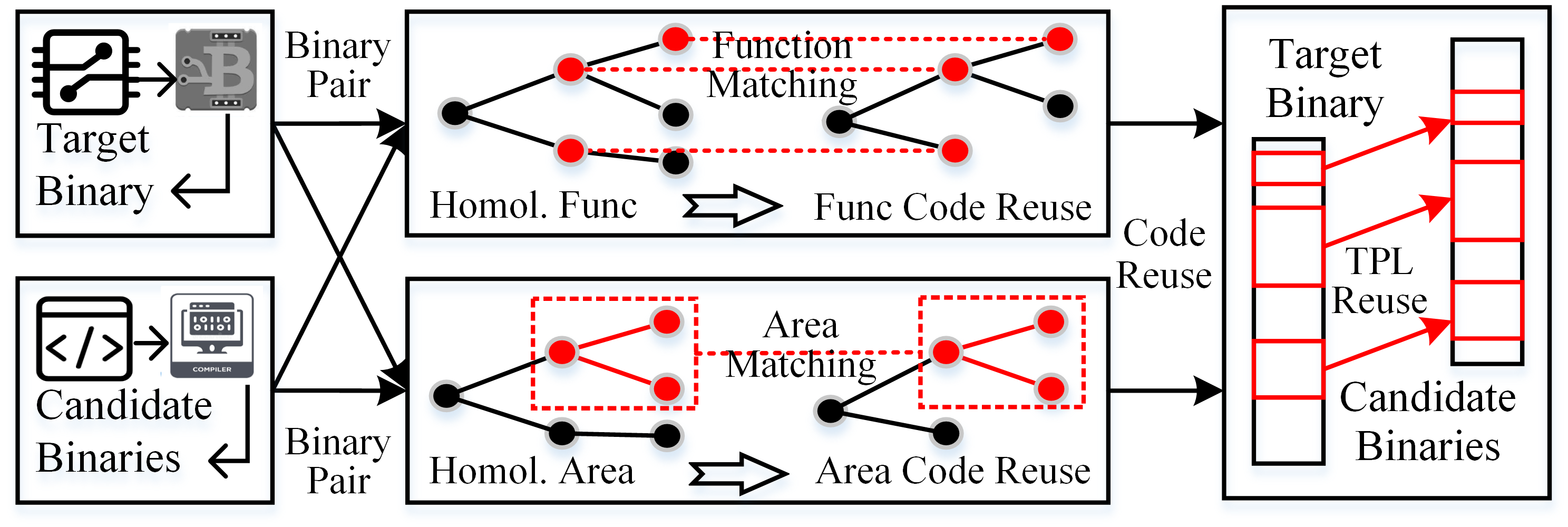}
	    \caption{Workflow of the Existing TPL Detection Framework. 
    The framework takes the binary pair as input, performs function or area matching to identify homologous code, and determines the presence of TPL reuse between the binaries.}
	\label{fig:Fig1_General_Workflow}
\end{figure}

TPL detection can be categorized into source-oriented and binary-oriented methods. As our work targets software binaries, where source code is often unavailable, we focus on the binary-oriented category, where methods are further divided into syntax-based and semantic-based approaches.
syntax-based methods—such as LibDX~\cite{Libdx}, B2SFinder~\cite{B2sfinder}, and OSSPolice~\cite{OSSPolice}—leverage low-level syntactic features (e.g., strings, constants, and integers) to enable static analysis, but their robustness degrades when facing cross-architecture differences or diverse compiler optimizations. In contrast, semantic-based approaches—such as MODX~\cite{ModX}, ISRD~\cite{ISRD}, and LibDB~\cite{Libdb}—improve resilience by capturing function-level semantic features through deep neural networks. 
Incorporating structural context, LibAM~\cite{Libam} further enhances detection accuracy by aligning matched code areas according to structural patterns in Function Call Graphs (FCGs).
Despite diverse feature representations, existing TPL detection methods generally follow a unified workflow (Fig.~\ref{fig:Fig1_General_Workflow}), in which reuse relationships are identified via function or area matching of the binaries.

\begin{figure}
	\centering
    \includegraphics[width=\columnwidth]{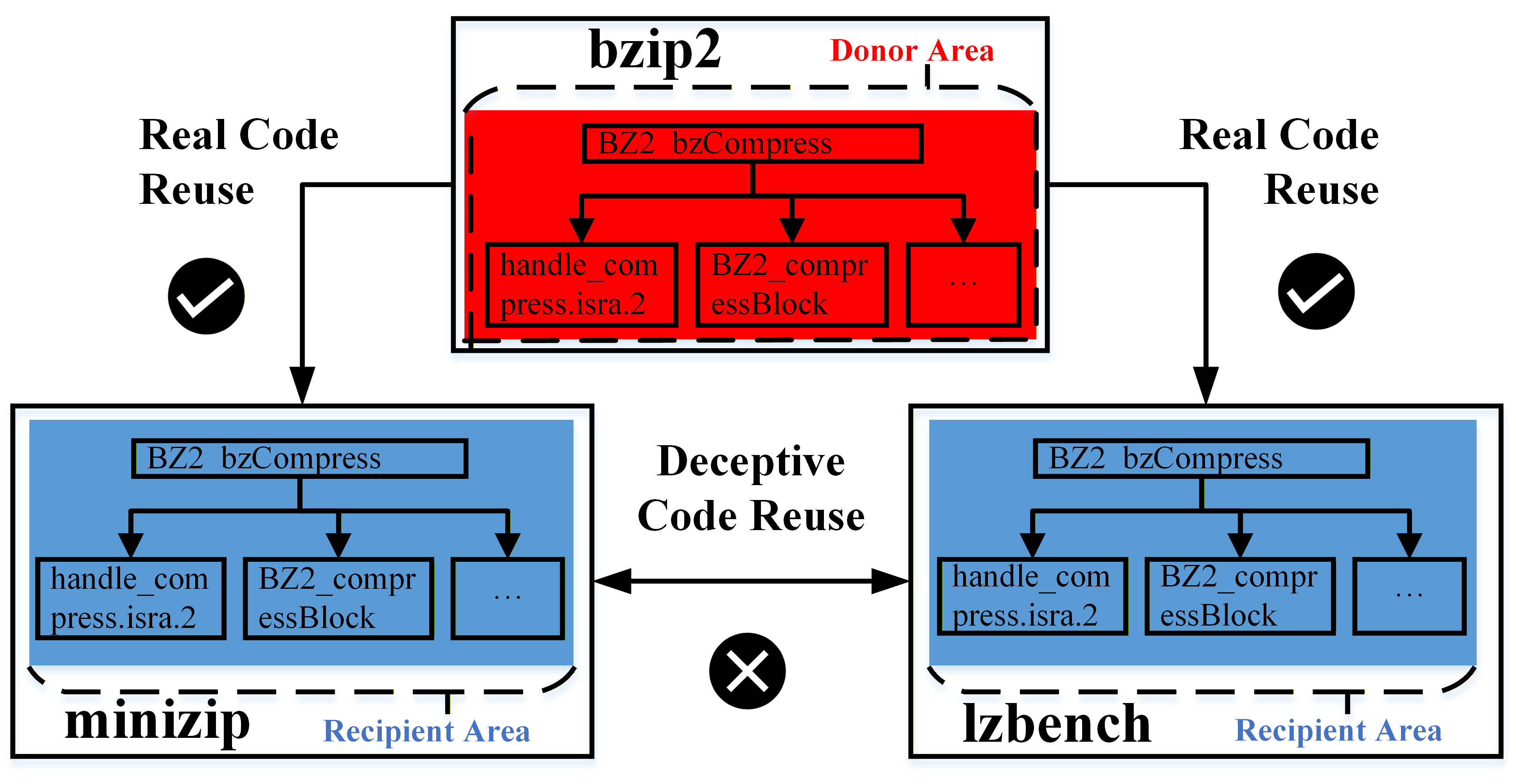}
	    \caption{Examples of Real and Deceptive Code Reuse. The reuse relationship between \texttt{bzip2} and \texttt{minizip} exemplifies real reuse, characterized by unidirectional code flow from donor to recipient, whereas the relationship between \texttt{minizip} and \texttt{lzbench} illustrates deceptive reuse, manifested as a bidirectional relationship without real code reuse.}
    \label{fig:Fig2_Example_Deceptive_Reuse}
\end{figure}

Existing TPL detection methods often neglect the directionality of code reuse, thereby conflating code matching with code reuse and ultimately leading to false-positive vulnerability reports, as illustrated in Fig.~\ref{fig:Fig2_Example_Deceptive_Reuse}.
In practice, real code reuse is defined as the unidirectional transfer of code from a TPL---acting as the provider of functional code---to a recipient code that incorporates the reused code.
In contrast, numerous cases have shown that two recipient code may independently incorporate the same functionality from a common TPL, thereby creating a bidirectional relationship between them, which is formally defined in this study as deceptive code reuse.
This type of reuse may falsely imply vulnerability propagation across unrelated projects, leading to spurious alerts in vulnerability analysis. Although LibDI~\cite{LibDI} advances the field by defining complex reuse scenarios, its methodology is limited by reliance on symbol information such as function names. A critical shortcoming is that it overlooks reuse directionality, thereby rendering LibDI unable to accurately distinguish between deceptive and real reuse. Consequently, establishing a mechanism to  differentiate between real and deceptive reuse within TPL detection represents an urgent research imperative.

To address these limitations, we propose \toolname{}, a three-stage TPL detection framework with the capability of deceptive reuse mitigation.
The first two stages, Function Matching and Area Alignment, are designed to identify homologous code areas between binaries, while the third stage, Reuse Discrimination, aims to determine the type of reuse between these code areas.
Acting as the preparatory phase, the Function Matching module applies Markov matrix encoding combined with MobileNet~\cite{Mobilenetv2} to identify homologous functions across binaries, while the Area Alignment module constructs Attributed Function Call Graphs (AFCGs) and employs a Graph Convolutional Network (GCN) to refine the alignment of \textbf{code areas}. The Reuse Discrimination stage forms the core of \toolname{}; it takes potential homologous code areas as input and models the donor--recipient relationships between them through two topology-sensitive features, thereby effectively distinguishing between real and deceptive reuse.
This design enables \toolname{} to distinguish real reuse from deceptive reuse, thereby ensuring precise TPL detection and reducing false-positive vulnerability reports caused by deceptive reuse.

We curated a dataset comprising deceptive reuse cases and used it to systematically evaluate \toolname{} against several state-of-the-art methods---including ISRD~\cite{ISRD}, LibDB~\cite{Libdb}, LibAM~\cite{Libam} and LibDI~\cite{LibDI}---in the tasks of reuse discrimination and TPL detection. 
Experimental results demonstrate that \toolname{} is highly effective in reuse discrimination, achieving an F1 score of 96.9\%. Furthermore, it attains an F1 score of 97.7\% in TPL detection, where it outperforms the best-performing baseline by 9.2\%.
Moreover, \toolname{} demonstrates robust performance across different architectures and compiler optimization levels, achieving average F1 scores of 92.9\% and 92.0\%, respectively.
Finally, a real-world case study demonstrates that \toolname{} outperforms the commercial tool BinaryAI ~\cite{BinaryAI} in mitigating deceptive reuse and enables more effective TPL and vulnerability detection.
We summarize our main contributions below:

\begin{itemize}
    \item We formally define deceptive reuse based on directionality to distinguish it from real reuse, illustrating with real-world examples its impact on false-positive vulnerability reports.
    \item We design algorithms to capture donor–recipient relationships between code areas and develop \toolname{}, a progressive framework for TPL detection that effectively mitigates deceptive reuse.
    \item \toolname{} achieves effective cross-architecture and cross-compilation optimization-level TPL detection, effectively mitigating deceptive reuse and outperforming the commercial tool BinaryAI.
\end{itemize}

\section{Preliminaries} \label{sec:preliminaries}
\subsection{TPL Detection Framework}

TPL detection frameworks typically identify reuse by analyzing binary pairs. As illustrated in Fig.~\ref{fig:Fig1_General_Workflow}, existing approaches generally utilize code matching at the function or area granularity to detect homologous code, which forms the basis of the detection process. To facilitate the subsequent discussion, we begin by defining key terminology.

\textbf{Binary pair}, denoted as $(B_{\mathrm{tgt}}, B_{\mathrm{cdd}})$, consists of two binaries. The target binary \(B_{\mathrm{tgt}}\) is extracted from a software package and serves as the object of TPL detection. In contrast, the candidate binary \(B_{\mathrm{cdd}}\) is compiled from the TPL source code and serves as the reference for TPL detection.

\textbf{Homologous code} is classified based on code granularity into homologous functions and homologous areas. The former denotes a pair of functions from $(B_{\mathrm{tgt}}, B_{\mathrm{cdd}})$ that originate from the same source, whereas the latter denotes a pair of code areas---each consisting of adjacent functions connected in the FCG---that derive from the same source.

In the TPL detection process, these frameworks identify homologous code between $(B_{\mathrm{tgt}}, B_{\mathrm{cdd}})$ through function or area matching. The matched homologous code serves as an indicator of code reuse, and TPL reuse between binaries is determined by evaluating whether the number of such matches exceeds a predefined threshold.

\begin{figure*}
	\centering
	\includegraphics[width=\textwidth]{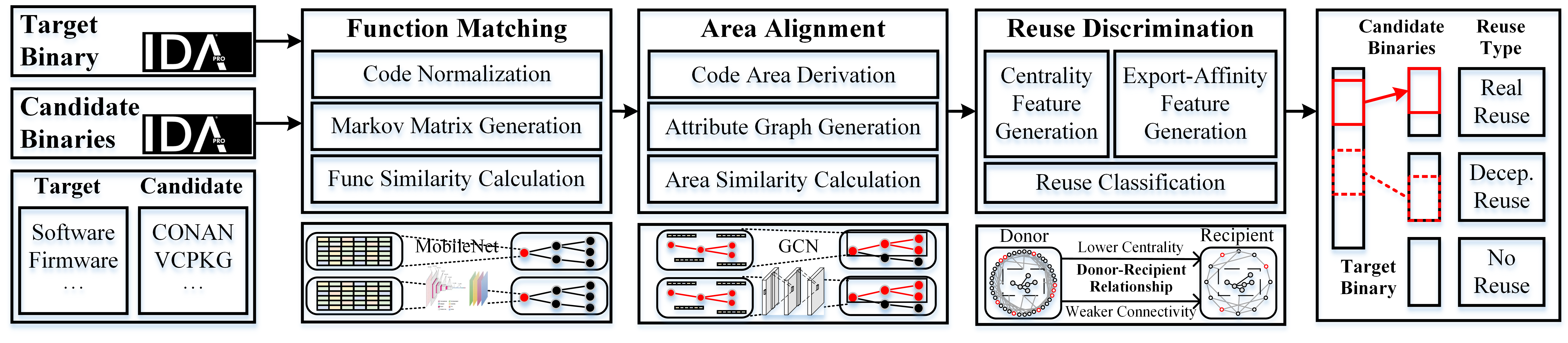}
	\caption{Workflow of the Three-Stage Binary Reuse Detection System \textbf{DERED}. The system comprises three stages: (1) Function Matching, which identifies homologous function pairs between target and candidate binaries; (2) Area Alignment, which establishes correspondences among these functions to delineate reused code areas; and (3) Reuse Discrimination, which categorizes each aligned area as real reuse, deceptive reuse, or no reuse.}
	  \label{fig:Fig3_Workflow_DeRed}
\end{figure*}

\subsection{Exposure of deceptive reuse} 

Existing TPL detection frameworks equate code matching with code reuse, overlooking the distinction between matching---a bidirectional relationship---and reuse---a unidirectional relationship.By extracting topology-sensitive features from $(B_{\mathrm{tgt}}, B_{\mathrm{cdd}})$ and modeling donor-recipient relationships between code areas based on their attributes, we can effectively distinguish between real and deceptive reuse.

\textbf{Reuse Discrimination and TPL Detection:} Reuse discrimination specifically distinguishes real from deceptive reuse within local code regions, whereas TPL detection determines whether a valid real reuse relationship exists at the binary level between $B_{\mathrm{tgt}}$ and $B_{\mathrm{cdd}}$.

\textbf{Donor and Recipient Areas:} 
Code areas within a $(B_{\mathrm{tgt}}, B_{\mathrm{cdd}})$ are classified as either donor or recipient based on their functional roles in the reuse chain, thereby facilitating reuse discrimination through the analysis of donor-recipient relationships. 
A Donor area provides reusable functionality to other binaries, as exemplified by \texttt{bzip2} (red box in Fig.~\ref{fig:Fig2_Example_Deceptive_Reuse}). Conversely, a Recipient area integrates code from a donor, as seen with \texttt{minizip} (blue box in Fig.~\ref{fig:Fig2_Example_Deceptive_Reuse}).

\textbf{Real and Deceptive Reuse}: 
Code reuse is classified by transfer direction. Real reuse occurs when a recipient code area directly incorporates donor code area, indicated by the unidirectional arrow from \texttt{bzip2} to \texttt{minizip} in Fig.~\ref{fig:Fig2_Example_Deceptive_Reuse}. Conversely, deceptive reuse arises between two recipients code area sharing a common donor, represented by the bidirectional arrow between \texttt{minizip} and \texttt{lzbench} in Fig.~\ref{fig:Fig2_Example_Deceptive_Reuse}.

\subsection{Detriment of deceptive reuse} 

The TPL-based vulnerability detection pipeline consists of three stages. First, it identifies TPL reuse to establish the relationship between \(B_{\mathrm{tgt}}\) and \(B_{\mathrm{cdd}}\). Second, it retrieves known vulnerability information corresponding to the identified \(B_{\mathrm{cdd}}\). Third, it propagates the vulnerability association to \(B_{\mathrm{tgt}}\) based on the inferred reuse relationship. 
Critically, the validity of vulnerability reporting relies on the accuracy of the initial detection phase. Consequently, this paper focuses on TPL detection. We contend that failing to differentiate deceptive reuse from real reuse inevitably leads to high false-positive rates in downstream vulnerability detection.

As illustrated in Fig.~\ref{fig:Fig2_Example_Deceptive_Reuse}, we denote \texttt{minizip} as \(B_{\mathrm{tgt}}\) and \texttt{bzip2}---which contains the vulnerable function \texttt{BZ2\_Vul}---as \(B_{\mathrm{cdd}}\). The unidirectional arrow represents a valid vulnerability association, indicating the correct propagation of vulnerability information. 
Consider the scenario where a bidirectional matching relationship exists between \texttt{minizip} and \texttt{lzbench}, as both components exhibit high similarity despite the absence of a valid unidirectional reuse relationship. Existing TPL detection methods, which predominantly rely on syntactic or semantic feature matching, overlook the inherent directionality of reuse detection, failing to effectively distinguish between real and deceptive reuse. Consequently, these tools incorrectly infer a reuse relationship, leading to erroneous vulnerability reports. We provide empirical evidence of this phenomenon through a detailed case study in Section~\ref{sec:case_of_study}.

In large open-source ecosystems, widely adopted software is often integrated into thousands of downstream projects, leading to a rapid proliferation of deceptive reuse relationships. These misleading associations substantially increase the rate of false-positive vulnerability reports, especially when relying on automated approaches such as TPL reuse detection. Notably, mitigating such false positives through manual verification is resource-intensive, delays security response, and erodes trust in the detection systems.

\section{Method} \label{sec:method}

\subsection{Overview}

\toolname{} takes as input the binary pair $(B_{\mathrm{tgt}}, B_{\mathrm{cdd}})$ and processes it through three dedicated modules---Function Matching, Area Alignment, and Reuse Discrimination---as illustrated in Fig.~\ref{fig:Fig3_Workflow_DeRed}, enabling effective TPL detection. In the Function Matching module, \toolname{} efficiently matches homologous function pairs between \(B_{\mathrm{tgt}}\) and \(B_{\mathrm{cdd}}\), establishing the foundation for subsequent analysis. In the Area Alignment module, these paired functions are linked to delineate potential reused code areas, which are then aligned to determine homologous areas. Binary pairs without any such areas are defined as exhibiting no reuse. In the Reuse Discrimination module, each aligned area is rigorously classified as either real reuse or deceptive reuse. A binary pair is ultimately labeled as real reuse if it contains more than three verified instances of real code reuse; otherwise, it is categorized as deceptive reuse.

Overall, \toolname{} employs a three-stage mechanism: it first matches homologous functions, then aligns the corresponding homologous areas, and finally assigns one of three labels---no reuse, deceptive reuse, or real reuse---to each aligned area. Ultimately, it determines the final TPL reuse relationship based on these discrimination labels.

\begin{table}[htbp]
    \centering
    \setlength{\tabcolsep}{12pt}
    \caption{Opcode Semantic Mapping Table}
    \label{tab:opcode_mapping}
    \renewcommand{\arraystretch}{1.25}
    \begin{tabularx}{\linewidth}{@{} l >{\raggedright\arraybackslash}X @{}}
        \toprule
        \textbf{Descriptor} & \textbf{Description} \\
        \midrule
        \textsc{Arith} ($s_1$) & Integer or floating-point computation \\
        \textsc{Xfer} ($s_2$)  & Register or memory data movement \\
        \textsc{Cmp} ($s_3$)   & Value comparison and flag setting \\
        \textsc{Logic} ($s_4$) & Bitwise logical manipulation \\
        \textsc{Shift} ($s_5$) & Bit-level shift or rotate \\
        \textsc{Jmp} ($s_6$)   & Direct jump without condition \\
        \textsc{Bcc} ($s_7$)   & Branch on condition evaluation \\
        \bottomrule
    \end{tabularx}
\end{table}

\subsection{Function Matching} \label{sec:function_matching}

This module matches homologous functions between \(B_{\text{tgt}}\) and \(B_{\text{cdd}}\).
To overcome the challenges of structural variations caused by compiler settings and the computational overhead of processing thousands of functions, this module incorporates code normalization, Markov matrix generation, and function similarity analysis. This approach ensures robust yet rapid matching, establishing an efficient foundation for subsequent analysis within \toolname{}.

\paragraph{\textbf{Code Normalization}}

This step normalizes assembly code to reduce compiler‑introduced noise—such as shifted address offsets, instruction variants (e.g., ``ADD'' vs. ``ADDS''), and cross‑architecture divergences (e.g., ``INC'' vs. ``ADD''). First, we extract the opcode sequence of function \(f\) as:

\begin{equation}
I_f = \{op_1,\,op_2,\,\ldots,op_n\}
\end{equation}

\noindent ordered by offset. Next, each opcode is mapped into one of the seven semantic categories as listed in Table~\ref{tab:opcode_mapping}, thereby producing a normalized code sequence of descriptors as:

\begin{equation}
D_f = \{d_1,\,d_2,\,\ldots, d_n\}
\end{equation}

\noindent where \(d_i\) is the descriptor for \(op_i\). Table~\ref{tab:opcode_mapping} lists each category and its description, enabling systematic normalization—for example, ``ADD'', ``ADDS'', and ``INC'' are all mapped to the arithmetic operator category, denoted as \(s_1\).

\paragraph{\textbf{Markov Matrix Generation}} 

The normalized code sequence \(D_f\) is used to construct a Markov matrix that encodes the key instruction transition patterns, thereby enabling efficient capture of function semantics. 
Specifically, a transition count matrix \(C \in \mathbb{N}^{7 \times 7}\) is constructed by iterating over adjacent instruction pairs \((d_t, d_{t+1})\) in the normalized sequence \(D_f\). Each entry \(c_{ij}\) captures the frequency of transitions from semantic descriptor \(s_i\) to \(s_j\), formally defined as:

\begin{equation}
c_{i,j} = \sum_{t=1}^{n-1} \delta(d_t = s_i) \cdot \delta(d_{t+1} = s_j)
\end{equation}

\noindent where $n$ is the length of $D_f$, $\delta(\cdot)$ is the indicator function, and $\mathcal{S} = \{s_1, \dots, s_7\}$ is the set of semantic descriptors.
By row-normalizing \(C\), we derive the Markov matrix \(P_f \in \mathbb{R}^{7 \times 7}\), where each entry \(p_{ij}\) is defined as:

\begin{equation}
P_f =  \{ p_{i,j} | p_{i,j} =\frac{c_{i,j}}{\sum_{t=1}^{7} c_{i,t}} \}
\end{equation}

\noindent The Markov matrix \( P_f \) captures two properties: structural compactness and semantic abstraction.
Structural compactness transforms variable-length sequences into fixed-dimensional vector representations, permitting end-to-end analysis by neural networks.
Semantic abstraction models normalized transition probabilities, thereby smoothing cross-architecture differences among homologous functions.

\paragraph{\textbf{Function Similarity Calculation}} 

To efficiently encode function-level Markov representations, we employ a MobileNetV2~\cite{Mobilenetv2} backbone, which strikes an effective balance between representational capacity and computational cost. 

The input matrix pair \((P_{f_a}, P_{f_b})\) is processed by a dual-branch MobileNetV2 backbone within a Siamese architecture, enabling the model to learn consistent representations through symmetric embedding pathways. Each branch incorporates hierarchical depthwise separable convolutions with skip connections, allowing the network to capture hierarchical dependencies within the transition distribution. This compact design supports efficient modeling of the Markov semantic structures. This compact and shared-parameter design supports efficient modeling of Markov transition distributions. 
Adopting this Markov representation offers distinct advantages over alternative function matching approaches. Specifically, this representation mitigates structural noise~\cite{Markov1} and models execution flow as spatial textures~\cite{Markov2}, thereby enabling the CNN backbone to extract topological features robust to compilation variations. Second, regarding computational efficiency, our matrix-based formulation circumvents the bottlenecks of alternative architectures. Unlike GNN-based methods~\cite{Gemini} limited by iterative message-passing or Transformer-based methods~\cite{NLP 1} constrained by quadratic self-attention, the Markov matrix leverages optimized matrix operations in conjunction with MobileNet to achieve superior efficiency.

We train the model to learn function embeddings by maximizing the similarity of homologous pairs while minimizing that of non-homologous ones. A function pair $(f_a, f_b)$ is then identified as homologous if the similarity score between their embeddings exceeds a predefined threshold (\(\theta_{func}\)=0.65). All identified pairs are subsequently forwarded to the Area Alignment module for further analysis. If no matches are detected, the binaries are classified as unrelated in terms of code reuse.

\subsection{Area Alignment} \label{sec:area_alignment}

This module aligns homologous areas between \(B_{\text{tgt}}\) and \(B_{\text{cdd}}\). To overcome the unreliability of traditional topology-based matching caused by optimization-induced structural variations, this module employs AFCG and a Siamese GCN architecture. By fusing stable semantic descriptors with neighborhood-aware graph convolutions, this approach effectively smooths out structural noise. This ensures resilient area alignment, establishing a precise foundation for subsequent analysis within \toolname{}.

\paragraph{\textbf{Code Area Derivation}}

This step derives an extended code area for each matched function by recursively traversing its reachable callees in the FCG to extract the FCG subgraph. The design is motivated by the observation~\cite{Libam} that code reuse patterns tend to cluster within localized call hierarchies centered on matched functions.

By expanding the call neighborhood of each matched function, the method  captures areas with a high likelihood of reuse and reliably preserves essential topological dependencies, thereby improving the overall effectiveness of area alignment. For each matched function pair \((f_{tgt}, f_{cdd})\), the procedure produces a corresponding pair of neighborhood subgraphs, denoted as \((g_{f_{tgt}}, g_{f_{cdd}})\), which serve as the reliable foundation for subsequent graph-based analysis.

\paragraph{\textbf{Attributed Graph Generation}}

This step converts each neighborhood subgraph \(g_f\) into an AFCG \(g_f^{attr}\) by representing function nodes as semantic embeddings derived from their corresponding Markov matrices. The design is grounded in the practical reality that each \( g_{f_{tgt}} \) is typically aligned against a large number of \( g_{f_{cdd}} \), and that the performance of purely topology-based matching tends to deteriorate as the candidate set scales up.

By encoding instruction-level behavioral patterns as semantic vectors and embedding them within the topology of the FCG, this step captures both local operational semantics and inter-function structural dependencies, enabling robust and distinctive area representations. 
For each neighborhood subgraph pair \((g_{f_{tgt}}, g_{f_{cdd}})\), this step generates a corresponding pair of AFCGs, denoted as \((g_{f_{tgt}}^{attr}, g_{f_{cdd}}^{attr})\), which are subsequently used as input for similarity computation.

\paragraph{\textbf{Area Similarity Calculation}}

This step encodes the AFCGs of code areas from both binaries and computes their similarity using a Siamese GCN architecture. Specifically, it takes AFCG pairs \((g_{f_a}^{attr}, g_{f_b}^{attr})\) as input and processes them with dual-channel GCNs. The symmetric structure enables joint representation learning of both input graphs.

Each channel of the network applies a stack of three GCN layers to extract both semantic and structural features from the input graph pairs. In general, a GCN layer updates the representation of each node by aggregating information from its local neighborhood, thereby capturing the graph’s topology and node attributes. Formally, in the \(l\)-th GCN layer, the representation of node \(i\), denoted as \(h_i^{(l)}\), is computed based on the representations of its adjacent nodes in the \((l-1)\)-th layer. This update is governed by the following rule:
 
\begin{equation}
h_i^{(l)} = \sigma\left(\sum_{j \in \mathcal{N}(i)} \frac{1}{c_{ij}} W^{(l-1)} h_j^{(l-1)} + b^{(l-1)}\right)
\end{equation}

\noindent where \(c_{ij}\) is a normalization constant that accounts for the degrees of the nodes \(i\) and \(j\), \(W^{(l-1)}\) is the learnable weight matrix, \(b^{(l-1)}\) is the bias vector in the \((l-1)\) -th layer, and \(\sigma(\cdot)\) denotes a non-linear activation function. Through repeated topology-aware aggregation and semantic transformation, node embeddings progressively capture structural patterns and semantic information.

We train a homologous area alignment model to learn area-level embeddings that maximize similarity for homologous areas and minimize it for non-homologous ones. Based on homologous function pairs identified by the function matching module, we construct AFCG pairs as alignment inputs. An area pair is labeled as aligned if their embedding similarity exceeds a predefined threshold (\(\theta_{area}\)=0.74). All aligned area pairs between \(B_{\mathrm{tgt}}\) and \(B_{\mathrm{cdd}}\) are then forwarded to the Reuse Discrimination module for further analysis; if no aligned areas are detected, the binaries are classified as unrelated with respect to reuse.

\subsection{Reuse Discrimination} \label{sec:reuse_discrimination}

\begin{figure}
	\centering
    \includegraphics[width=\columnwidth]{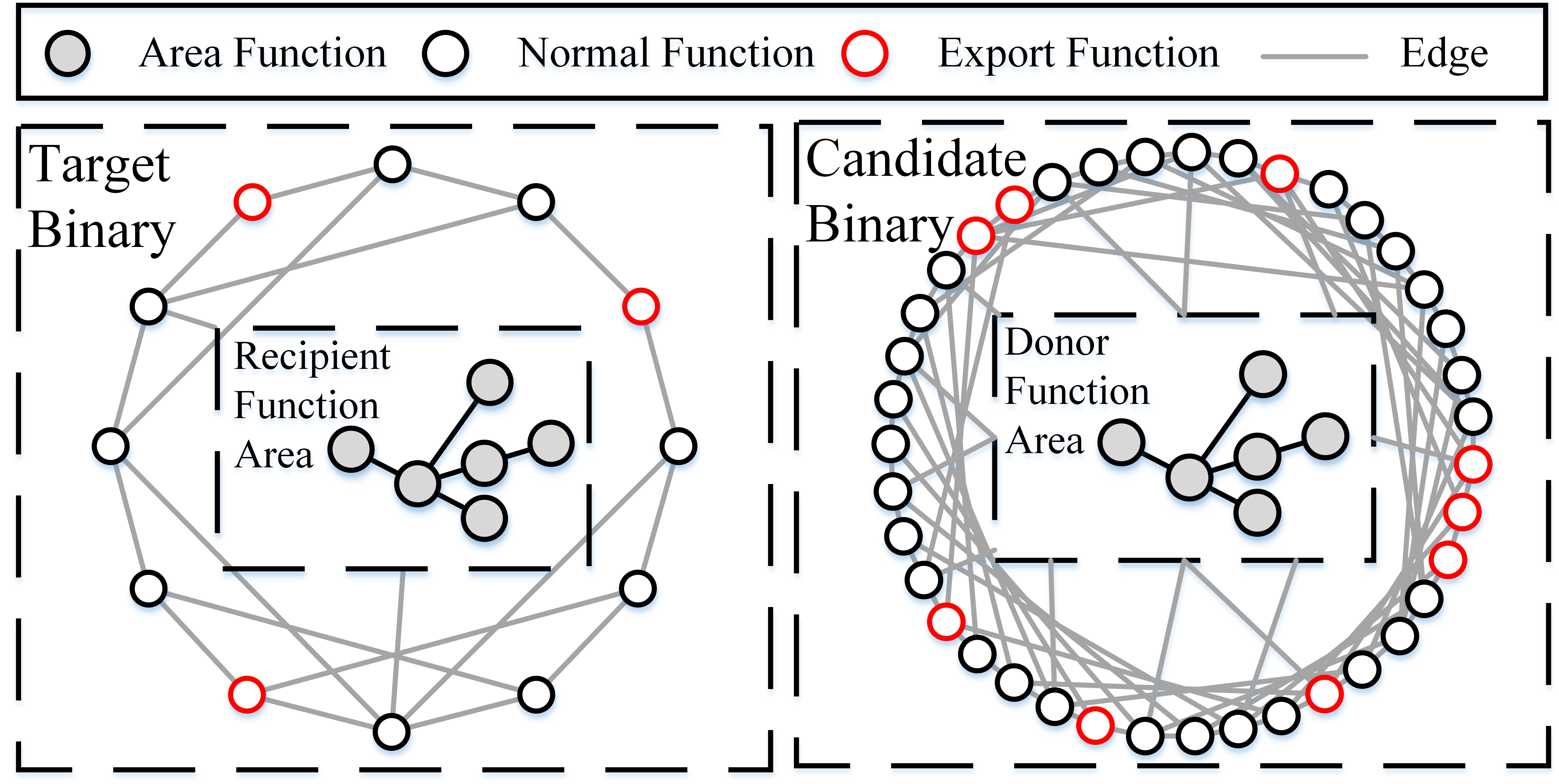}
	    \caption{Topological Differences Between Donor and Recipient Areas. Donor areas in their respective binaries exhibit higher centrality and stronger connectivity to exported functions, whereas Recipient areas present simpler topological structures with substantially fewer interconnections.}
    \label{fig:Fig4_Topology_Diff}
\end{figure}

This module differentiates donor areas from recipient areas by analyzing the topological differences between homologous areas and designing two algorithms to extract topology-sensitive features. Building upon these features, it further employs advanced ensemble learning techniques to construct a model of the relationships between homologous area pairs, thereby enabling the effective discrimination between real reuse and deceptive reuse.

As shown in Fig.~\ref{fig:Fig4_Topology_Diff}, donor areas encapsulate core functionalities within the \(B_{\mathrm{cdd}}\), exhibiting stronger centrality and export-linked characteristics. In contrast, recipient areas contribute auxiliary logic to the \(B_{\mathrm{tgt}}\), displaying simpler structures with fewer such connections. Building on this observation, we perform an in-depth analysis of these differences and introduce two topologically-sensitive features: a centrality feature, which evaluates the structural prominence of a code area within the binary based on its topological connectivity and functional impact; and an export-affinity feature, which measures the strength of its association with the binary’s exported functions. 
In the following sections, we will detail the algorithms used to extract these two features and explain how they can be applied to facilitate reuse discrimination and TPL detection.

\paragraph{\textbf{Centrality Score Calculation}}

Existing work applies graph-based centrality measures to assess node importance in dependency graphs~\cite{TPLite}. Extending this approach to binary analysis, we employ the PageRank algorithm to compute centrality scores for aligned area pairs and introduce interaction weights to reduce the effect of non-interactive centrality, leading to more effective evaluation.

We begin with the conventional PageRank algorithm, which estimates the relative importance of each function within the binary’s FCG through an iterative process. This algorithm provides a probabilistic assessment of a function’s centrality within its local neighborhood. For each function \(f_i\) in a neighborhood subgraph \(g_{f}\) (as defined in Section~\ref{sec:area_alignment}), its influence at iteration \(t\) is computed as:

\begin{equation}
    \text{PR}(f_i)^{(t)} = \frac{1 - d}{N} + d \sum_{f_j \in B(f_i)} \frac{\text{PR}(f_j)^{(t-1)}}{Call(f_j)}
\end{equation}

\noindent where \(N\) is the total number of functions in the code area, \(d\) is the damping factor, \(B(f_i)\) denotes the set of functions that invoke \(f_i\), and \(Call(f_j)\) represents the number of outgoing calls from function \(f_j\). While the PageRank score reflects the global structural prominence of functions in the binary, it does not fully capture the contextual interactions between a function within the area and external functions. To address this limitation, we introduce an interaction weight \(\omega(f_i)\) for each function \(f_i\), quantifying its external connectivity relative to its overall interactions. This weight is defined as:

\begin{equation}
\omega(f_i) = \frac{|E_{\text{ext}}(f_i)|}{|E_{\text{int}}(f_i)| + |E_{\text{ext}}(f_i)|},
\end{equation}

\noindent where \(E_{\text{int}}(f_i)\) and \(E_{\text{ext}}(f_i)\) denote the sets of intra-area and inter-area function call edges. Finally, the customized centrality score of function \(f_i\) is computed by weighting its PageRank score with \(\omega(f_i)\). Accordingly, the overall centrality score of an area \(g\) is given by the average weighted importance of all its constituent functions:

\begin{equation}
C(g) = \frac{1}{|g|} \sum_{f_i \in g} \omega(f_i) \cdot \text{PR}(f_i)
\end{equation}

\noindent where \(|g|\) is the number of functions in area \(g\). For each area pair \((g_{f_a}, g_{f_b})\), we compute their respective centrality features \(C(g_{f_a})\) and \(C(g_{f_b})\), which are subsequently used as input to the reuse discrimination model.

\paragraph{\textbf{Export-Affinity Score Calculation}}

Existing work identifies relationships between code areas by directly counting the number of exported functions between two areas, which lacks robustness~\cite{LibDI}. Extending this approach to achieve a deeper understanding of their interconnections, we extract transition paths from exported functions to associated areas within the binary, compute cumulative transition probabilities along all such paths, and define the export-affinity score as the average Cumulative Transition Probability (CTP).

We begin with a reverse reachability analysis that identifies all valid execution paths from each exported function $e_j \in E = \{e_1, e_2, \dots, e_n\}$ to the area $g$. A valid transition path $\pi$ is defined as a sequence of function calls:

\begin{align}
\pi &= (e_j \to f_1 \to f_2 \to \dots \to f_t) \notag \\
&\quad \text{where } f_t \in g,\ \forall i < t,\ f_i \notin g
\end{align}

\noindent where $f_t \in g$, and for all preceding nodes $\forall i < t,\ f_i \notin g$. This constraint guarantees that each path originates strictly from outside the target area and enters $g$ only at the terminal function. We aggregate all such paths from $e_j$ to $g$ into a set $P_{e_j \to g}$, which serves as the foundation for the subsequent structural affinity estimation. Next, we quantify the structural reachability of the area from each exported function by computing the CTP for every path in $P_{e_j \to g}$. 
Given the absence of dynamic profiling data, we adopt a uniform branching assumption. By distributing probability mass uniformly across outgoing edges, this assumption induces a structural decay effect that inherently penalizes paths with high fan-out. This design ensures that the affinity score measures structural tightness rather than execution likelihood; consequently, direct, quasi-deterministic calls contribute significantly more than connections mediated by complex, multi-branch structures. We mathematically operationalize this logic as a probabilistic chain.
For a given path $\pi$, its CTP is formally defined as the product of edge-level transition probabilities along the path:

\begin{equation}
\text{CTP}(\pi) = \prod_{(f_i \to f_{i+1}) \in \pi} \frac{1}{|\mathrm{succ}(f_i)|}
\end{equation}

\noindent where $\mathrm{succ}(f_i)$ denotes the set of immediate callees of function $f_i$. This formulation assumes equal probability for all branches from a given function node in the function call graph. The total CTP for $e_j$ with respect to area $g$ is computed by summing over all valid paths:

\begin{equation}
\text{CTP}^{\text{total}}(e_j, g) = \sum_{\pi \in P_{e_j \to g}} \text{CTP}(\pi)
\end{equation}

\noindent which provides a probabilistic measure of how strongly $e_j$ interacts with $g$ through all feasible call paths. Finally, we aggregate the CTP scores of all exported functions toward area $g$ to compute the average CTP, and assess the overall export-affinity of the area:

\begin{equation}
E(g) = \frac{1}{|E|} \sum_{e_j \in E} \text{CTP}^{total}(e_j,g)
\end{equation}

\noindent where $|E|$ denotes the number of exported functions in the binary. The score $E(g)$ quantifies the interaction intensity between the exported interface and the code area, as determined by the call graph topology. For each area pair $(g_{f_a}, g_{f_b})$, we derive the export-affinity features $E(g_{f_a})$ and $E(g_{f_b})$, which are used as inputs to the reuse discrimination model to identify code reuse patterns. It should be noted that if $|E|=0$, then $E(g)$ is defined to be zero, ensuring the mathematical consistency of the formulation and preventing undefined behavior in cases where no exported functions are present in the binary.

\paragraph{\textbf{Area Classification}}

This step integrates the centrality and export-affinity scores of code areas in both the \(B_{\mathrm{cdd}}\) and \(B_{\mathrm{tgt}}\), enabling effective discrimination between real and deceptive code reuse.
For each code area, the following four topologically-sensitive features are extracted:

\begin{itemize}
\item \(C(g_{f_a})\): Centrality Score of \(g_{f_a}\) in \(B_{tgt}\)
\item \(E(g_{f_a})\): Export-affinity Score of \(g_{f_a}\) in \(B_{tgt}\)
\item \(C(g_{f_b})\): Centrality Score of \(g_{f_b}\) in \(B_{cdd}\)
\item \(E(g_{f_b})\): Export-affinity Score of \(g_{f_b}\) in \(B_{cdd}\)
\end{itemize}

These features jointly capture the internal importance of an area—reflected by centrality—and its external interaction potential—quantified by export-affinity—both essential for reuse analysis. 
To enable reliable classification based on these features, we adopt the gradient boosting framework LightGBM~\cite{LightGBM}, which efficiently handles complex interactions, scales better than traditional machine learning techniques, and requires significantly less computational power compared to deep learning methods.
The reuse discrimination model can be formalized as follows:

\begin{equation}
\text{RT} = \text{LightGBM}(C(g_{f_a}), E(g_{f_a}), C(g_{f_b}), E(g_{f_b}))
\end{equation}

\noindent where \(\text{RT} \in \{0,1\}\) represents the reuse type classification label, with 1 indicating real reuse between homologous areas, and 0 indicating deceptive reuse. All homologous area pairs exhibiting real reuse between \(B_{\text{tgt}}\) and \(B_{\text{cdd}}\) are subsequently used for final TPL detection. If more than three homologous area with real reuse are identified, the binaries are classified as exhibiting real reuse; otherwise, they are classified as exhibiting deceptive reuse.

\section{Experiments} \label{sec:experiments}

In this section, we present a comprehensive evaluation of the proposed \toolname{} by comparing its effectiveness with a range of baselines across diverse tasks. The experiments are designed to address the following research questions:

\begin{itemize}
    \item \textbf{RQ1:} How effectively does \toolname{} mitigate deceptive reuse, and how does this improve the effectiveness of TPL detection?
    \item \textbf{RQ2:} How do topologically-sensitive (i.e., Centrality Feature and Export-Affinity Feature) features improve the effectiveness of TPL detection in \toolname{}?
    \item \textbf{RQ3:} How do function matching and area alignment enhance TPL detection effectiveness of \toolname{}?
    \item \textbf{RQ4:} How well does \toolname{} maintain the robustness of TPL detection across diverse architectures and compiler optimization levels?
\end{itemize}

\subsection{Experiment Settings \& Metrics}

The experimental system is built on Ubuntu 22.04, with hardware including an Intel Xeon 128-core 3.0 GHz processor (with hyper-threading support), 1 TB of memory, and dual NVIDIA V100 32 GB GPUs, providing ample resources for algorithm execution and comparison. Notably, due to its lightweight nature, \toolname{} can also run effectively on systems without GPU support. For binary disassembly, we use the industry-standard decompiler IDA Pro 9.1 for reverse engineering and file analysis, whose robust feature set meets the evaluation requirements. The core system is implemented in Python 3.8, leveraging Python's extensive library ecosystem while ensuring engineering convenience.
Our framework trains three distinct models: MobileNetv2 for function matching, a GCN for area alignment, and LightGBM for reuse discrimination. Regarding training strategies, the deep learning models (MobileNetv2 and GCN) were optimized using the Adam optimizer with a weight decay of $5 \times 10^{-4}$ and an initial learning rate of 0.001. We configured the batch size to 128 for MobileNetv2 and 256 for the GCN. Both models underwent training for a maximum of 100 epochs, employing Early Stopping (patience=10) and a dropout rate of 0.5 to prevent overfitting. Conversely, the LightGBM classifier was configured with 100 estimators, a learning rate of 0.1, and 31 leaves. All datasets were randomly partitioned into training, validation, and testing sets with a ratio of 8:1:1.

\begin{figure}
	\centering
    \includegraphics[width=\columnwidth]{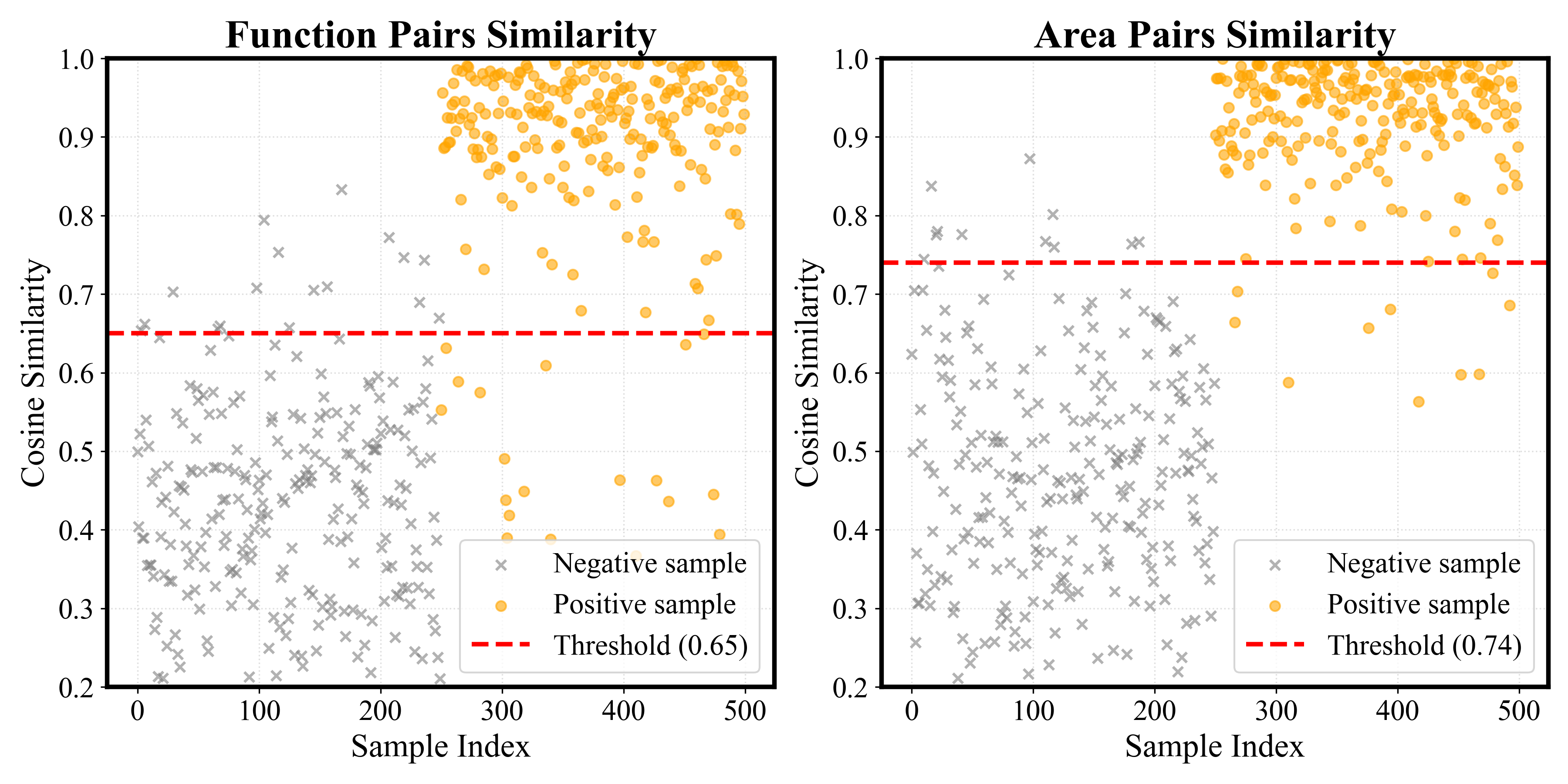}
	    \caption{Cosine Similarity Distributions for Function and Area Pairs. Red dashed lines mark empirical thresholds of 0.65 and 0.74, effectively separating positive (orange dots) and negative (gray crosses) samples. This distinct demarcation validates the robustness of the thresholds in identifying homologous code.}
    \label{fig:Fig5_Threshold_Plot}
\end{figure}

To determine the optimal decision boundaries, we employed an empirical calibration strategy based on the training partitions of the D-Func and D-Area datasets. We analyzed the probability density functions of cosine similarities for positive and negative samples within these training sets to identify intersection points that maximize class separation. This process yielded thresholds of \(\theta_{func}\)=0.65 and \(\theta_{area}\)=0.74, respectively.
To verify the generalization capability of these empirically derived values, we visualized the distribution of a representative subset of 500 pairs randomly sampled from the held-out validation sets (unseen during threshold selection). This subset size was selected specifically to ensure visual clarity and avoid overplotting, allowing for a distinct observation of the distribution trends. As illustrated in Fig.~\ref{fig:Fig5_Threshold_Plot}, the selected thresholds align with the natural demarcation between homologous and non-homologous samples in the validation data, confirming their robustness without overfitting to the training distribution.

To evaluate \toolname{}, we use standard metrics such as Precision, Recall, F1 Score , and the Area Under the Receiver Operating Characteristic Curve (AUC-ROC). These metrics are applied across tasks such as function matching, area alignment, reuse discrimination, and TPL detection. Precision measures the proportion of correctly predicted positive instances among all predicted positives, reflecting the accuracy of positive predictions. Recall measures the proportion of actual positive instances that are correctly identified, capturing the completeness of detection. The F1 Score is the harmonic mean of Precision and Recall, providing a balanced evaluation that accounts for both false positives and false negatives. Furthermore, the ROC curve illustrates the trade-off between the True Positive Rate (TPR) and the False Positive Rate (FPR), with the AUC quantifying the model's overall capability to distinguish between classes across various thresholds.

\begin{table*}[t]
    \centering
    \caption{Comparison of Key Features between \textbf{DERED} and Baseline Methods.}
    \label{tab:baseline_comparison}
    \small
    \renewcommand{\arraystretch}{1.4}
    
    \begin{tabularx}{\textwidth}{l P{4.2cm} L l}
        \toprule
        \textbf{Method} & \textbf{Function Matching} & \textbf{Area Alignment Mechanism} & \textbf{Reuse Discrimination} \\
        \midrule
        
        \textbf{ISRD}~\cite{ISRD} & 
        \multirow{4}{=}{%
            \textbf{Graph-based (Gemini)}~\cite{Gemini}: Structure2Vec embedding recursively aggregating block attributes over ACFG. 
            \newline
            \textbf{code-based (Jtrans)}~\cite{NLP 1}: Jump-aware BERT encoding control flow dependencies.
        } & 
        \textbf{ISRD$_{area}$: }Matched function ratio $> 50\%$. & 
        \multirow{3}{3.75cm}[-0.5em]{\RaggedRight No reuse discrimination mechanism; simple code matching is directly regarded as code reuse.}\\
        
        \cmidrule(r){1-1} \cmidrule(lr){3-3}
        
        \textbf{LibDB}~\cite{Libdb} & & 
        \textbf{LibDB$_{area}$: }Connected subgraph size $\ge 3$. & 
        \\ 
        
        \cmidrule(r){1-1} \cmidrule(lr){3-3}
        
        \textbf{LibAM}~\cite{Libam} & & 
        \multirow{2}{=}{%
            \textbf{LibAM/LibDI$_{area}$: }Structural Similarity encoded via Siamese GNN and Alignment Length Similarity of connected areas.
        } & 
        \\ 
        
        \cmidrule(r){1-1} \cmidrule(l){4-4}
        
        \textbf{LibDI}~\cite{LibDI} & & & 
        Text-Based Disc. \\
        
        \midrule
        \addlinespace[0.5ex]
        
        \textbf{DERED} & 
        \textbf{DERED$_{func}$: }Markov CNN. & 
        \textbf{DERED$_{area}$: }GCN on AFCG structures. & 
        Topology-Based Disc. \\
        
        \bottomrule
    \end{tabularx}
\end{table*}

\subsection{Dataset}

To comprehensively evaluate the effectiveness and robustness of \toolname{} across various TPL reuse scenarios, including deceptive cases, we have curated three datasets: \textbf{D-Func}, \textbf{D-Area}, and \textbf{D-TPL}.

\textbf{D-Func}: 
For training and evaluation of the function matching model in \toolname{}, we utilize a benchmark dataset introduced in ~\cite{BCSD_review} ,which has been widely adopted in prior research. This dataset encompasses 7 representative projects, resulting in 24 distinct libraries compiled across three architectures (ARM, x86, x64) and four optimization levels (O0, O1, O2, O3). In total, the dataset comprises 2,689 binary files and 13.8 million functions, thereby offering diverse and realistic scenarios for model development. 

\textbf{D-Area}:
For the training and evaluation of the area alignment model in \toolname{}, we employ a dataset derived from LibAM~\cite{Libam}, a well-established and widely-recognized dataset in the field of TPL detection. This dataset includes 289 binary files from 24 popular projects, spanning three architectures and four optimization levels. The dataset contains a total of 758,880 homologous areas. LibAM’s widespread adoption in benchmarking stems from its diverse and representative binary collection, ensuring a rigorous evaluation of the proposed model.

\textbf{D-TPL}: 
For the training and evaluation of the reuse discrimination and TPL detection model in \toolname{}, we first crawled and analyzed the dependency metadata of 1,864 projects from the Conan repository—which encompasses inter-TPL reuse relationships such as \texttt{minizip} reusing \texttt{zlib} and \texttt{bzip2}—and constructed a dictionary to capture their reuse relationships, formatted as \texttt{\{`minizip': [`zlib', `bzip2']\}}. 
Informed by this dictionary, we curated a set of 83 representative projects for compilation, specifically targeting those involved in more than three reuse relationships, acting as either a donor or a recipient. This selection criterion ensures that the dataset captures complex and diverse reuse patterns prevalent in real-world software ecosystems.
Finally, using the Conan client in a Linux environment, we compiled the selected projects, generating 572 corresponding binaries, as a single project typically yields multiple binary artifacts. To facilitate a comprehensive evaluation of \toolname{}’s performance in TPL detection, reuse discrimination, and realistic scenario adaptation, we further constructed three dataset subsets based on the aforementioned binaries:

\begin{itemize}
    \item \textbf{D-TPL-1:} To evaluate \toolname{} in the TPL detection task, we identified real reuse samples directly from explicit dependency lineages in the Conan repository~\cite{Conan} and constructed deceptive reuse samples by pairing projects that share overlapping function names but lack dependency connections. To ensure label reliability, three inspectors independently conducted a comprehensive manual review of the entire candidate set. Disagreements were resolved via voting, ultimately yielding 300 pairs of verified genuine reuse and 288 pairs of deceptive reuse.

    \item \textbf{D-TPL-2:} To evaluate \toolname{} in the TPL detection task, we identified real reuse samples via direct function name matching. For deceptive reuse, we employed a two-phase pipeline: (1) Candidate Generation, identifying potential pairs by matching function names and structural features across unrelated projects; and (2) Hybrid Verification, utilizing DeepSeek-V2~\cite{deepseek} with chain-of-thought prompting to analyze semantic divergence. The prompting strategy directs the model to evaluate reuse directionality and authenticity by examining package naming conventions, copyright declarations, and directory hierarchies.  The final filtered dataset comprises 30,000 genuine and 30,000 deceptive reuse pairs. 

    \item \textbf{D-TPL-3:} cross-architecture and cross-optimization TPL detection tasks, we utilized the binaries compiled under three architectures (\texttt{x86}, \texttt{x64}, and \texttt{ARM}) and four optimization levels (\texttt{O0}, \texttt{O1}, \texttt{O2}, and \texttt{O3}). Following the same generation pipeline and validation criteria as D-TPL-1, we extended the scope to include these diverse compilation settings. Based on the established reuse relationships, we randomly selected 10,000 binary pairs exhibiting real reuse and 10,000 pairs exhibiting deceptive reuse under cross-compilation settings.
\end{itemize}

\subsection{Baseline}

To comprehensively evaluate the effectiveness of \toolname{} across function matching, area alignment, reuse detection, and TPL detection, Table~\ref{tab:baseline_comparison} systematically summarizes the key technical differences across these three stages.

First, regarding function matching, \toolname{} employs a custom Markov matrix approach combined with a lightweight CNN, as detailed in \S\ref{sec:function_matching}. In contrast, to ensure a fair comparison, we standardized the matching step for all baselines using Gemini~\cite{Gemini} and JTrans~\cite{NLP 1}, representing state-of-the-art graph-based and code-based embedding approaches, respectively. Subsequently, \toolname{} and the baselines perform area alignment based on these matched functions. Specifically, \toolname{} conducts alignment using GCNs on AFCG structures (\S\ref{sec:area_alignment}), whereas the baseline methods adopt the following distinct mechanisms:
\begin{itemize}
    \item \textbf{ISRD~\cite{ISRD}}: Adopts a ratio-based criterion, confirming area alignment when more than 50\% of the functions in an area are matched based on similarity.
    \item \textbf{LibDB~\cite{Libdb}}: Enforces topological constraints, requiring at least three  functions form a connected subgraph within the area’s FCG to establish alignment.
    \item \textbf{LibAM~\cite{Libam}}: Determines alignment by grouping matched functions into connected components and evaluating them using a hybrid metric that combines structural similarity with alignment length similarity.
    \item \textbf{LibDI~\cite{LibDI}}: Inherits the alignment strategy of LibAM but extends it by incorporating an additional reuse discrimination mechanism, as detailed below.
\end{itemize}

\noindent Finally, regarding \textbf{reuse discrimination}, only \toolname{} and \textbf{LibDI} perform this specific verification stage. While \toolname{} leverages topology-sensitive features of function areas (\S\ref{sec:reuse_discrimination}), LibDI performs reuse discrimination by analyzing string constants and exported function names.

\subsection{Answering to RQ1}

\begin{table}[t]
\caption{Comparative Evaluation of \textbf{DERED} and Baselines for Reuse Discrimination and TPL Detection (\textbf{RQ1}). }
\begin{center}
\begin{tabularx}{\columnwidth}{c|c|*{3}{>{\centering\arraybackslash}X}}
\toprule
\textbf{Exp.} & \textbf{Method} & \textbf{Precision} & \textbf{Recall} & \textbf{F1} \\
\midrule
\multirow{5}{*}{\shortstack{\textit{Reuse} \\ \textit{Discrim} \\ \textit{(Area)}}}
& $\text{ISRD}$~\cite{ISRD} & $16.5\%$ & $74.9\%$ & $27.0\%$ \\
& $\text{LibDB}$~\cite{Libdb} & $46.3\%$ & $79.6\%$ & $58.5\%$ \\
& $\text{LibAM}$~\cite{Libam} & $72.7\%$ & $86.1\%$ & $78.8\%$ \\
& $\text{LibDI}$~\cite{LibDI} & $82.1\%$ & $88.4\%$ & $85.1\%$ \\
& \toolname{} & $\mathbf{96.2\%}$ & $\mathbf{97.8\%}$ & $\mathbf{96.9\%}$ \\
\midrule
\multirow{5}{*}{\shortstack{\textit{TPL} \\ \textit{Detect} \\ \textit{(Binary)}}}
& $\text{ISRD}$~\cite{ISRD} & $19.4\%$ & $70.3\%$ & $30.4\%$ \\
& $\text{LibDB}$~\cite{Libdb} & $44.1\%$ & $81.6\%$ & $57.2\%$ \\
& $\text{LibAM}$~\cite{Libam} & $76.8\%$ & $91.2\%$ & $83.4\%$ \\
& $\text{LibDI}$~\cite{LibDI} & $84.6\%$ & $92.8\%$ & $88.5\%$ \\
& \toolname{} & $\mathbf{97.2\%}$ & $\mathbf{98.3\%}$ & $\mathbf{97.7\%}$ \\
\bottomrule
\end{tabularx}
\label{tab:TAB_RQ1_1}
\end{center}
\end{table}

To address RQ1, we design two interrelated experiments focusing on reuse discrimination and TPL detection.
The first experiment (\textit{Reuse Discrim}) aims to prove that \toolname{} can effectively mitigate deceptive reuse.
The second experiment (\textit{TPL Detect}) aims to prove that \toolname{} can enhance TPL detection effectiveness by mitigating deceptive reuse.
In both experiments, performance is measured with three standard metrics—precision, recall, and F1 score—and \toolname{} is evaluated against four representative baselines: ISRD, LibDB, LibAM, and LibDI.

\textit{Reuse Discrim} evaluates the effectiveness of each method in reuse discrimination at the code area level, processing area pairs labeled as real or deceptive reuse from the \textbf{D-TPL-2} dataset. 
As demonstrate in Table~\ref{tab:TAB_RQ1_1}, \toolname{} achieves a precision of 96.2\%, a recall of 97.8\%, and an F1 score of 96.9\%, indicating a clear improvement in precision and a substantial reduction in false positives compared to baselines.
However, \toolname{} still exhibits a small number of false positives caused by ambiguous donor-recipient relationships, particularly in minute code fragments where topological distinctiveness is low.
Among the baselines, LibDI performs the best, with a precision of 82.1\%, a recall of 88.4\%, and an F1 score of 85.1\%. LibAM follows with 72.7\% precision, 86.1\% recall, and 78.8\% F1 score, while LibDB shows moderate performance and ISRD exhibits the weakest overall performance. 
Compared to LibDI, \toolname{} shows absolute improvements of 14.1\% in precision, 9.4\% in recall, and 11.8\% in F1 score, highlighting its enhanced effectiveness in mitigating deceptive reuse.

\begin{figure}
	\centering
    \includegraphics[width=\columnwidth]{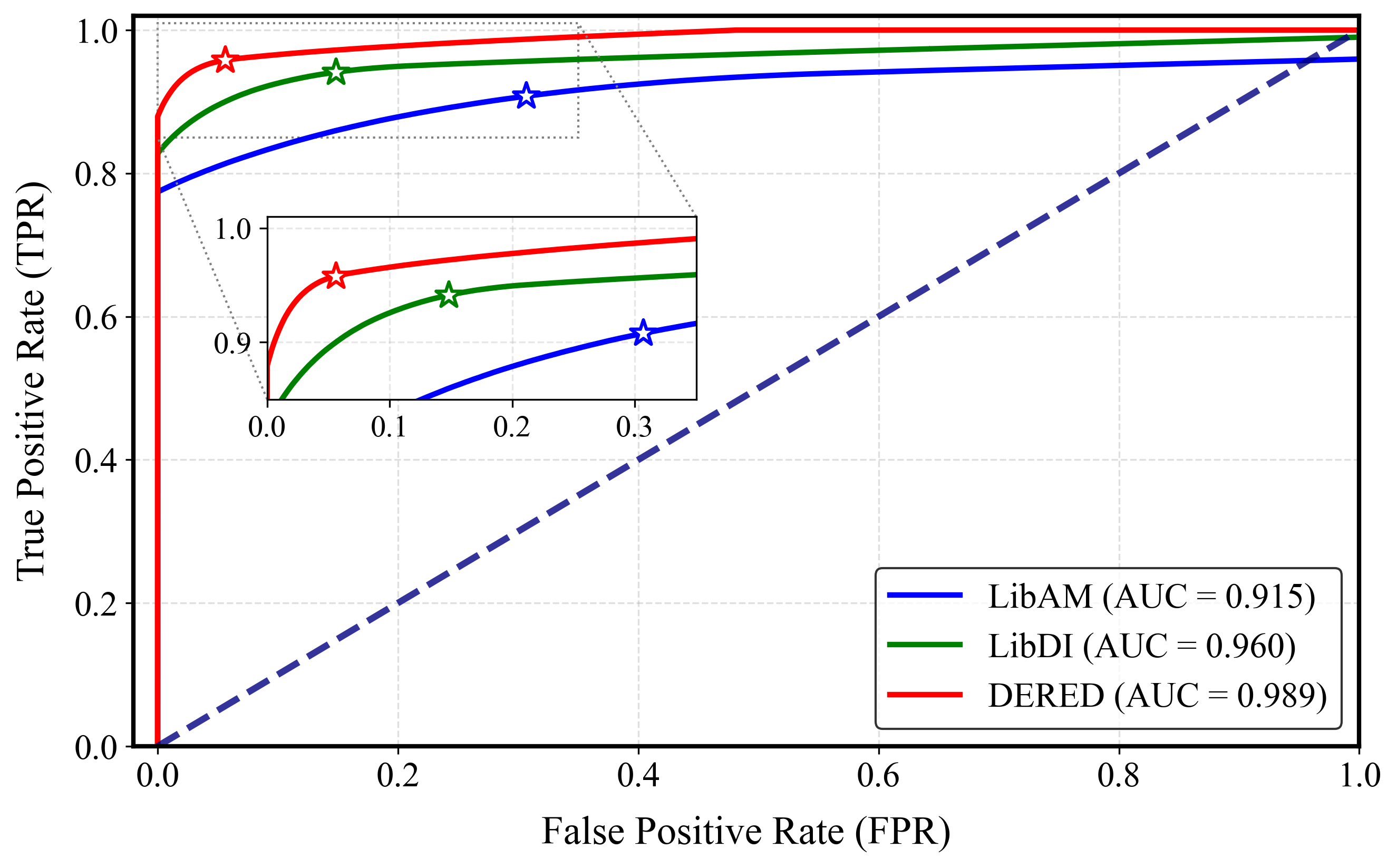}
    \caption{ROC Curves Evaluating TPL Detection Effectiveness. \textbf{DERED} (red, AUC=0.989) outperforms LibDI (green, 0.960) and LibAM (blue, 0.915). The zoomed-in inset highlights \textbf{DERED}'s rapid ascent at low FPR thresholds.}
    \label{fig:Fig6_ROC_CURVE}
\end{figure}

\textit{TPL Detect} evaluates the effectiveness of each method in TPL detection at the binary level, processing binary pairs labeled as real or deceptive reuse from the \textbf{D-TPL-1} dataset. 
As demonstrate in Table~\ref{tab:TAB_RQ1_1}, \toolname{} achieves a precision of 97.2\%, a recall of 98.3\%, and an F1 score of 97.7\%, indicating that the mitigation of deceptive reuse enables \toolname{} to identify more reliable reuse areas and thereby improve overall detection effectiveness.
Among the baselines, LibDI performs the best, with a precision of 84.6\%, a recall of 92.8\%, and an F1 score of 88.5\%. LibAM follows with 76.8\% precision, 91.2\% recall, and 83.4\% F1 score, while LibDB shows moderate performance and ISRD exhibits the weakest overall performance.
Compared with LibDI, \toolname{} improves precision by 12.6\%, recall by 5.5\%, and F1 score by 9.2\%, confirming that the mitigation of deceptive reuse contributes to reducing false positives and improving effectiveness in TPL detection.
To evaluate robustness beyond point metrics, Figure~\ref{fig:Fig6_ROC_CURVE} presents the ROC curves for \toolname{}, LibDI, and LibAM. \toolname{} achieves a dominant performance with an AUC of 0.989, surpassing LibDI and LibAM. Geometrically, the \toolname{} curve converges closely to the ideal coordinate, indicating high sensitivity with minimal error. Furthermore, the zoomed-in inset highlights that \toolname{} maintains a rapid ascent in TPR even at extremely low FPR thresholds. This trajectory confirms \toolname{} effectively suppresses false alarms, offering a superior sensitivity-specificity trade-off compared to baselines.

In summary, \toolname{} outperforms all baseline methods across the three evaluation metrics in both the \textit{Reuse Discrim} and \textit{TPL Detect} experiments, with especially pronounced improvements in precision.
These results confirm that the reuse discrimination module integrated into \toolname{}, which leverages topologically sensitive features to model the direction of code area reuse, effectively mitigates deceptive reuse, thereby providing more effective TPL detection.
In contrast, ISRD and LibDB, which rely on function-level code matching for TPL detection, fail to capture complex reuse relationships. Consequently, their overall performance across all metrics remains relatively low.
LibAM and LibDI both adopt area alignment strategies, which enable more accurate code matching compared to ISRD and LibDB, but they still fail to effectively mitigate deceptive reuse. Specifically, LibAM lacks a reuse discrimination mechanism, leading to more false positives and lower precision compared to LibDI. LibDI performs reuse discrimination through string and exported function features, which lack sufficient robustness, leading to an overall performance that remains inferior to that of \toolname{}.

\begin{center}
\begin{tcolorbox}[colframe=black, colback=white!10, colbacktitle=gray!30, coltitle=black, sharp corners=southwest, arc=4mm, boxrule=0.8mm, width=0.485\textwidth, title=\textbf{Answering to RQ1}]
\toolname{} leverages a reuse discrimination module to effectively mitigate deceptive reuse, thereby enabling more reliable TPL detection.
\end{tcolorbox}
\end{center}

\subsection{Answering to RQ2}

To address Research Question 2 (RQ2), we propose four variants of \toolname{}: \text{No-Topo}, \text{Central-Only}, \text{Export-Only}, and \text{Full-Model}. These variants are designed to facilitate ablation experiments on \toolname{}.
In \text{No-Topo}, topological snesitive features are excluded from both reuse discrimination and TPL detection; in \text{Central-Only}, only centrality features are incorporated; in \text{Export-Only}, only export-affinity features are considered; and in \text{Full-Model}, the complete implementation of \toolname{} is employed. All variants are evaluated under the same experimental setup as in RQ1.

\begin{table}[t]
\caption{Comparative Evaluation of Different Variants of \textbf{DERED} for Reuse Discrimination and TPL Detection (\textbf{RQ2}).}
\begin{center}
\begin{tabularx}{\columnwidth}{c|c|*{3}{>{\centering\arraybackslash}X}}
\toprule
\textbf{Exp.} & \textbf{Variant} & \textbf{Precision} & \textbf{Recall} & \textbf{F1} \\
\midrule
\multirow{4}{*}{\shortstack{\textit{Reuse} \\ \textit{Discrim.} \\ \textit{(Area)}}}
& $\text{No-Topo}$ & $75.8\%$ & $87.1\%$ & $81.1\%$ \\
& $\text{Central-Only}$ & $87.5\%$ & $93.7\%$ & $90.5\%$ \\
& $\text{Export-Only}$ & $83.1\%$ & $92.4\%$ & $87.3\%$ \\
& $\textbf{Full-Model}$  & $\mathbf{96.2\%}$ & $\mathbf{97.8\%}$ & $\mathbf{96.9\%}$ \\
\midrule
\multirow{4}{*}{\shortstack{\textit{TPL} \\ \textit{Detect.} \\ \textit{(Binary)}}}
& $\text{No-Topo}$ & $80.7\%$ & $91.9\%$ & $85.9\%$ \\
& $\text{Central-Only}$ & $89.5\%$ & $95.3\%$ & $92.3\%$ \\
& $\text{Export-Only}$ & $85.1\%$ & $94.1\%$ & $89.3\%$ \\
& $\textbf{Full-Model}$ & $\mathbf{97.2\%}$ & $\mathbf{98.3\%}$ & $\mathbf{97.7\%}$ \\
\bottomrule
\end{tabularx}
\label{tab:TAB_RQ2_1}
\end{center}
\end{table}

As demonstrate in Table~\ref{tab:TAB_RQ2_1}, \toolname{} exhibits substantial performance variation under different variants of \toolname{}.  
Using the \text{No-Topo} configuration, the model achieves only $75.8\%$ precision, $87.1\%$ recall, and an $81.1\%$ F1 score in reuse discrimination, and $80.7\%$ precision, $91.9\%$ recall, and an $85.9\%$ F1 score in TPL detection.  
When the \text{Central-Only} configuration is applied, the results improve considerably, reaching $87.5\%$ precision, $93.7\%$ recall, and a $90.5\%$ F1 score in reuse discrimination, alongside $89.5\%$ precision, $95.3\%$ recall, and a $92.3\%$ F1 score in TPL detection.  
In contrast, the \text{Export-Only} configuration yields slightly weaker outcomes compared with \text{Central-Only}, achieving $83.1\%$ precision, $92.4\%$ recall, and an $87.3\%$ F1 score in reuse discrimination, and $85.1\%$ precision, $94.1\%$ recall, and an $89.3\%$ F1 score in TPL detection.  
Most notably, the \text{Full-Model} achieves significantly better results---$96.2\%$ precision, $97.8\%$ recall, and a $96.9\%$ F1 score in reuse discrimination, and $97.2\%$ precision, $98.3\%$ recall, and a $97.7\%$ F1 score in TPL detection---representing absolute improvements of $11.8\%$, $8.4\%$, and $5.4\%$ in F1 score over the No-Topo, Export-Only, and Central-Only variants, respectively, marking the indispensability of integrating topological features for both effective reuse discrimination and reliable TPL detection.

In summary, \toolname{} demonstrates that the inferior effectiveness of individual feature arises from their limited ability to capture the nuanced patterns of reuse type. 
Specifically, Centrality-based features focus on the global structure, effectively identifying the significance of code areas within the entire binary; however, they may fail to account for the dependency relationships within local interactions. On the other hand, Export-based features emphasize the interactions between code areas and externally exported functions by analyzing local paths, yet they are insufficient in capturing the intricate dependencies within the global structure. 
The integration of Centrality and Export-Affinity features effectively captures both global and local characteristics, thus mitigating the limitations of individual features and offering a more holistic understanding of reuse patterns.

\begin{center}
\begin{tcolorbox}[colframe=black, colback=white!10, colbacktitle=gray!30, coltitle=black, sharp corners=southwest, arc=4mm, boxrule=0.8mm, width=0.485\textwidth, title=\textbf{Answering to RQ2}]
$\text{\toolname{}}$ integrates topology-sensitive features, providing a comprehensive understanding of reuse patterns, enabling effective TPL detection.
\end{tcolorbox}
\end{center}

\subsection{Answering to RQ3}

\begin{table}[t]
\caption{Comparative Evaluation of \textbf{DERED} and Baselines for Function Matching and Area Alignment (\textbf{RQ3})}
\begin{center}
\begin{tabularx}{\columnwidth}{c|c|*{4}{>{\centering\arraybackslash}X}}
\toprule
\textbf{Exp.} & \textbf{Method} & \textbf{Prec.} & \textbf{Rec.} & \textbf{F1} & \textbf{Time} \\
\midrule
\multirow{3}{*}{\shortstack{\textit{Func} \\ \textit{Match}}}
& $\text{Gemini}$~\cite{Gemini} & $95.2\%$ & $93.8\%$ & $94.5\%$ & $0.183\,\text{ms}$ \\
& $\text{Jtrans}$~\cite{NLP 1} & $\mathbf{96.7\%}$ & $\mathbf{95.8\%}$ & $\mathbf{96.3\%}$ & $0.736\,\text{ms}$ \\
& \textbf{DERED}$_{func}$ & $93.5\%$ & $91.7\%$ & $92.6\%$  & $\mathbf{0.011}\,\textbf{ms}$ \\
\midrule
\multirow{4}{*}{\shortstack{\textit{Area} \\ \textit{Align}}}
& $\text{ISRD}$~\cite{ISRD} & $38.6\%$ & $67.3\%$ & $49.1\%$ & $0.023\,\text{ms}$ \\
& $\text{LibDB}$~\cite{Libdb} &  $71.5\%$ & $85.3\%$ & $77.8\%$ & $0.054\,\text{ms}$ \\
& $\text{LibAM}$~\cite{Libam} &  $\mathbf{95.8\%}$ & $93.1\%$ & $94.4\%$ & $0.169\,\text{ms}$ \\
& \textbf{DERED}$_{area}$ & $95.2\%$ & $\mathbf{96.3\%}$ & $\mathbf{95.8\%}$ & $0.144\,\text{ms}$ \\
\bottomrule
\end{tabularx}
\label{tab:TAB_RQ3_1}
\end{center}
\end{table}

To address RQ3, we performed a cross-methodological evaluation of \toolname{} using the \textbf{D-Func} and \textbf{D-Area} datasets. For the \textit{Func Match} task, we employed Gemini~\cite{Gemini} and JTrans~\cite{NLP 1} as representative baselines for graph-based and code-based paradigms, respectively. For \textit{Area Align}, we benchmarked against ISRD~\cite{ISRD}, LibDB~\cite{Libdb}, and LibAM~\cite{Libam}. Crucially, in addition to the accuracy metrics utilized in previous RQs, this study evaluates computational overhead, specifically the execution time required for analyzing individual functions and areas.

For \textit{Func Match}, the results in Table~\ref{tab:TAB_RQ3_1} show that while baselines like JTrans and Gemini yield F1 scores of 96.3\% and 94.5\% respectively, \toolname{} maintains a competitive accuracy of 92.6\%. In terms of efficiency, \toolname{} requires only 0.011 ms per function, representing a speedup of approximately 67 times over JTrans (benchmarked with a batch size of 128). Analytically, these figures highlight a strategic trade-off: despite a marginal 3.7\% difference in F1 score compared to JTrans, \toolname{} achieves orders-of-magnitude improvements in speed. This efficiency stems from our Markov-based formulation, which avoids the computationally expensive feature extraction inherent in complex graph-based or code-based baselines.

For \textit{Area Align}, the results in Table~\ref{tab:TAB_RQ3_1} show that \toolname{} achieves a superior F1-score of 95.8\%. Notably, compared to the leading baseline LibAM, \toolname{} improves accuracy by 1.4\% while reducing inference time by approximately 14.8\%. In contrast, while baselines such as ISRD and LibDB exhibit high efficiency, they yield suboptimal F1 scores. These results demonstrate that \toolname{} effectively overcomes the accuracy-latency trade-off: unlike methods with limited representation capabilities (ISRD/LibDB), our Siamese-based GCN architecture efficiently captures hybrid code and structural semantic features, outperforming LibAM in both precision and speed.

In summary, considering both function matching and area alignment, \toolname{} improves efficiency while ensuring performance, making it the optimal method currently available.
For function matching, \toolname{} employs a Markov graph combined with a lightweight MobileNet, providing a significant efficiency advantage over traditional graph-based methods. In the area alignment phase, it utilizes an AFCG in conjunction with a Siamese-based GCN architecture, which not only ensures effectiveness but also offers higher efficiency compared to LibAM.

\begin{center}
\begin{tcolorbox}[colframe=black, colback=white!10, colbacktitle=gray!30, coltitle=black, sharp corners=southwest, arc=4mm, boxrule=0.8mm, width=0.485\textwidth, title=\textbf{Answering to RQ3}]
The Function Matching and Area Alignment modules of \toolname{} achieve a balanced trade-off between efficiency and performance, providing a solid foundation for reuse discrimination and TPL detection.
\end{tcolorbox}
\end{center}

\subsection{Answering to RQ4}

To address RQ4, we conducted a comprehensive evaluation of the TPL detection effectiveness of \toolname{} using the realistic scenarios subset \textbf{D-TPL-3} of the \textbf{D-TPL} dataset. The evaluation employed the F1 score as the metric and was carried out across four distinct settings: (1) \textit{NATIVE}, where the target and candidate binaries share the same architecture and optimization level; (2) \textit{CROSS\_ARCH}, where the binaries differ in architecture but share the same optimization level; (3) \textit{CROSS\_OPTI}, where the binaries share the same architecture but differ in optimization levels; and (4) \textit{CROSS\_BOTH}, where the binaries differ in both architecture and optimization levels.

\begin{figure}
	\centering
    \includegraphics[width=\columnwidth]{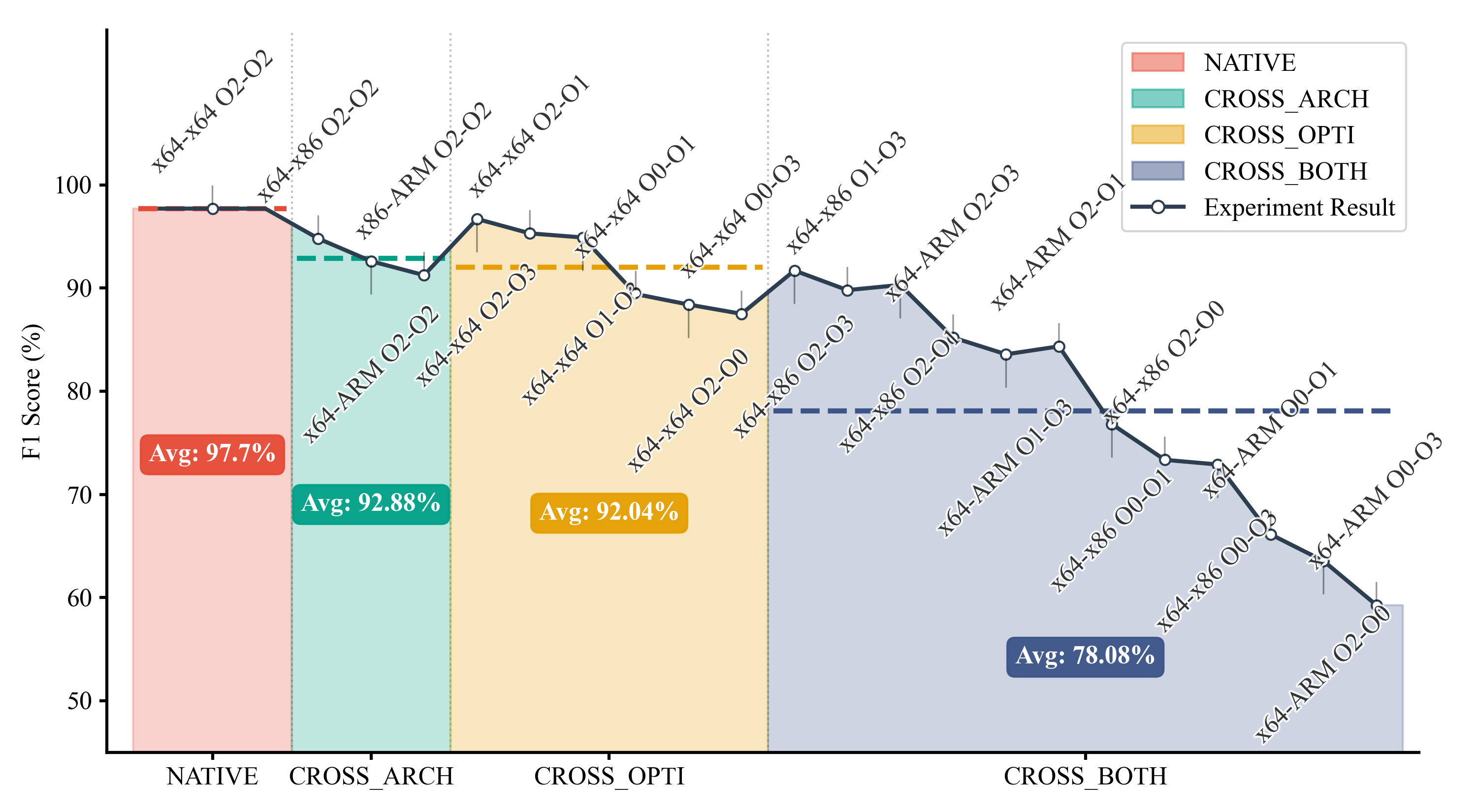}
    \caption{Evaluation of \textbf{DERED}'s effectiveness across four compilation scenarios: \textit{NATIVE}, \textit{CROSS\_ARCH}, \textit{CROSS\_OPTI}, and \textit{CROSS\_BOTH}. The markers connected by the solid line represent the F1 Scores for individual binary pairs, whereas the colored background regions denote the overall average performance for each scenario.}
    \label{fig:Fig7_Cross_Compilation}
\end{figure}

As shown in Fig.~\ref{fig:Fig7_Cross_Compilation}, the F1 scores for individual binary pairs are visualized using a solid line, while the overall average performance is highlighted by colored background regions. The results indicate that \toolname{} maintains strong TPL detection effectiveness in the majority of compilation scenarios.
In the \textit{NATIVE} setting, where binaries share the same architecture and optimization level, \toolname{} achieves the highest F1 score of 97.7\%, indicating its optimal performance under ideal consistency. 
In the \textit{CROSS\_ARCH} setting, where binaries differ in architecture, the F1 score decreases to an average of 92.9\%. 
In the \textit{CROSS\_OPTI} setting, where the architecture remains consistent but optimization levels vary, the average F1 score is 92.0\%. 
Finally, in the \textit{CROSS\_BOTH} setting, which involves simultaneous changes in both architecture and optimization, \toolname{} yields an average F1 score of 78.1\%.
This performance degradation is primarily attributed to significant code divergence resulting from the compounding effects of architectural heterogeneity and optimization disparities. 
These findings demonstrate that \toolname{} maintains robust performance across the \textit{CROSS\_ARCH}, \textit{CROSS\_OPTI}, and \textit{CROSS\_BOTH} scenarios. In all scenarios, it retains a reasonable level of detection capability.

In summary, \toolname{} demonstrates strong robustness in cross-architecture and cross-optimization TPL detection scenarios, which is attributed to the specialized design of its core modules.
The function matching module captures functional behavior patterns by constructing a Markov matrix from normalized opcode sequences, reducing reliance on specific instruction sets or optimization strategies. The area alignment module leverages the AFCG to model function call relationships and attribute features, alleviating inconsistencies caused by architectural variations and compiler optimizations. Finally, the reuse discrimination module detects code reuse through topologically sensitive features that remain stable across compilation scenarios, further strengthening the robustness of TPL detection.

\begin{center}
\begin{tcolorbox}[colframe=black, colback=white!10, colbacktitle=gray!30, coltitle=black, sharp corners=southwest, arc=4mm, boxrule=0.8mm, width=0.485\textwidth, title=\textbf{Answering to RQ4}]
Through the specialized design of its three core modules, \toolname{} exhibits strong robustness in TPL detection across diverse architectures and optimization levels, consistently maintaining high effectiveness even when both factors vary simultaneously.
\end{tcolorbox}
\end{center}

\section{Case Study} \label{sec:case_of_study}

To empirically validate the practical efficacy of \toolname{}, we conducted a comparative case study against BinaryAI~\cite{BinaryAI}, a representative commercial software composition analysis tool. While BinaryAI serves as a critical industry baseline, its inclusion in our large-scale quantitative evaluation was precluded by its closed-source nature, which enforces strict daily query quotas and anti-automation mechanisms. Consequently, we utilize this targeted case study as a representative qualitative evaluation. This approach allows for a rigorous manual verification of \toolname{} against industry standards, bypassing the limitations of batch-processing commercial APIs. 
In this experiment, \texttt{lzbench} was designated as the \(B_{\mathrm{tgt}}\), while \texttt{bzip2-1.0.6} (CVE-2019-12900) and \texttt{minizip-4.0.2} (CVE-2023-48107) served as the \(B_{\mathrm{cdd}}\). 
As illustrated in Fig.~\ref{fig:Fig2_Example_Deceptive_Reuse}, \texttt{lzbench} exhibits real code reuse from \texttt{bzip2}, confirming the presence of logic related to CVE-2019-12900. In contrast, the reuse associated with \texttt{minizip} is deceptive, meaning the vulnerable code for CVE-2023-48107 is physically absent from \texttt{lzbench}. 
Experimental results confirmed that BinaryAI effectively detected real reuse between \texttt{lzbench} and \texttt{bzip2}, but failed to filter deceptive reuse between \texttt{lzbench} and \texttt{minizip}, incorrectly flagging CVE-2023-48107 in \texttt{lzbench}.
This false positive arises because BinaryAI conflates code matching with code reuse and lacks the capability to identify deceptive reuse, which leads to the propagation of erroneous vulnerability reports.

\begin{figure}
	\centering
    \includegraphics[width=.94\columnwidth]{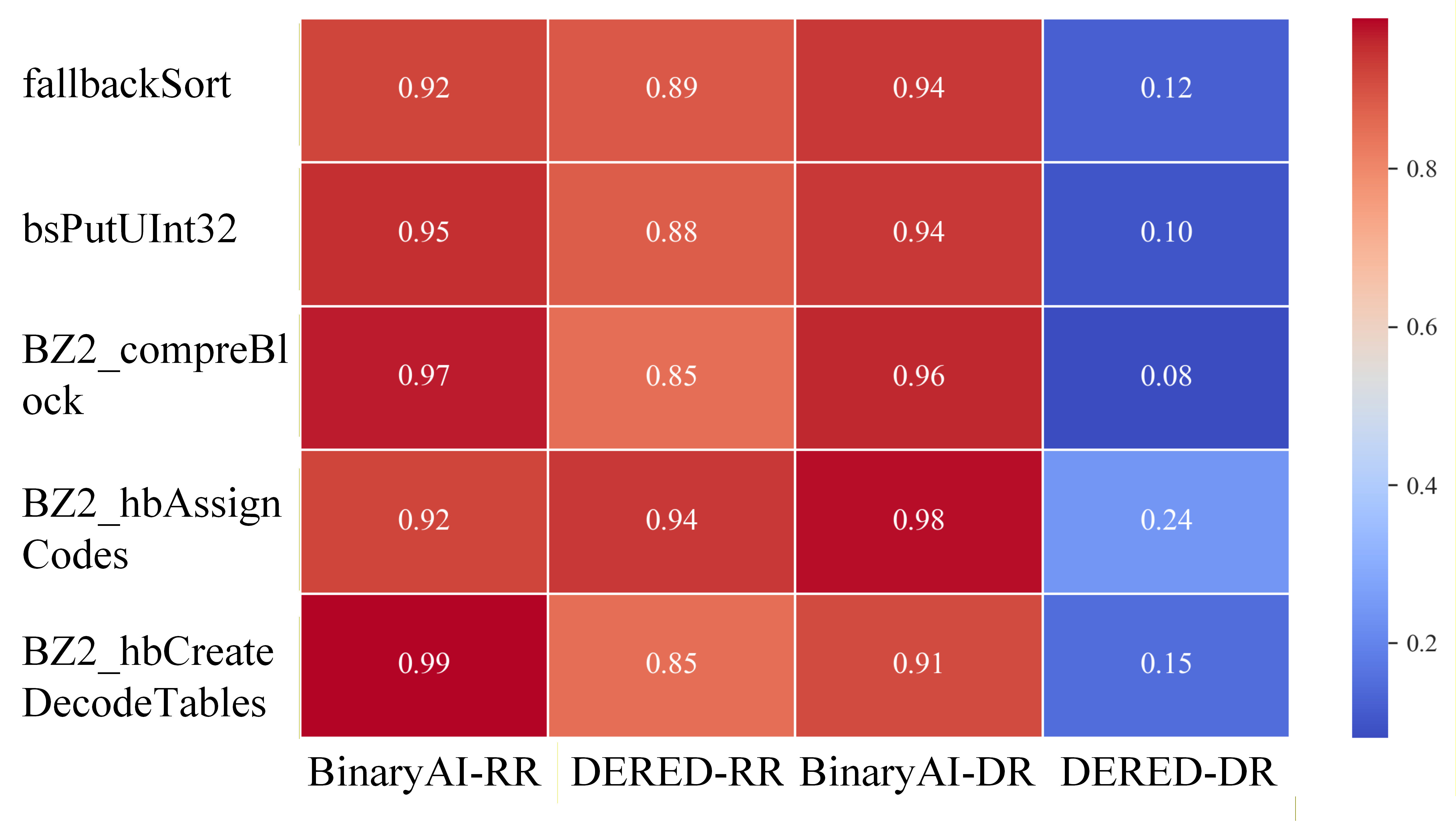}
	    \caption{Reuse Detection Performance of BinaryAI and \textbf{DERED} across five representative code areas. BinaryAI-RR and \textbf{DERED}-RR denote (real) reuse detection result between bzip2 and lzbench, while BinaryAI-DR and \textbf{DERED}-DR display (deceptive) reuse detection result between lzbench and minizip.}
    \label{fig:Fig8_Case_Study}
\end{figure}

As shown in Fig.~\ref{fig:Fig8_Case_Study}, we conducted a detailed comparison of code reuse detection using BinaryAI and \toolname{} on five specific code areas. 
The figure presents matching scores for each tool across real and deceptive code reuse scenarios. 
In real reuse scenarios, both tools demonstrate effective detection, with BinaryAI scoring between 0.92 and 0.99 and \toolname{} scoring between 0.85 and 0.94, indicating reliable identification of real reuse. However, a significant divergence is observed in deceptive scenarios. \toolname{} effectively distinguishes deceptive reuse by analyzing donor-recipient relationships, resulting in appropriately low matching scores of 0.08–0.15. Conversely, BinaryAI fails to differentiate deceptive cases and erroneously assigns high similarity scores ranging from 0.91 to 0.98.
Based on the experimental results, we clarified the mechanism of erroneous vulnerability propagation caused by deceptive reuse. Specifically, the code match between \texttt{lzbench} and \texttt{minizip} stems solely from the fact that both TPLs reuse \texttt{bzip2}. \toolname{} correctly identifies this indirect relationship and avoids establishing a non-existent link. In contrast, BinaryAI misidentifies this shared code as \texttt{lzbench} directly reusing \texttt{minizip}. Consequently, BinaryAI erroneously reports that \texttt{lzbench} contains CVE-2023-48107—a vulnerability specific to \texttt{minizip}. Manual verification confirmed that the vulnerable code corresponding to CVE-2023-48107 is physically absent from \texttt{lzbench}. Although such false positives could theoretically be eliminated through manual analysis, the prohibitive cost of intervention highlights the critical need for \toolname{}'s automated detection capabilities.

Leveraging the real reuse identified by \toolname{}, we further detected vulnerabilities in \texttt{lzbench}.
Specifically, \toolname{} located the vulnerable function $\textit{BZ2\_decompress}$  from \texttt{bzip2} within \texttt{lzbench}.
CVE-2019-12900 affects \texttt{bzip2} up to version~1.0.6 and was remediated in version~1.0.7. 
Since TPL detection typically lacks version-level granularity, we employed string analysis to confirm that the target binary corresponds to \texttt{lzbench 2.0.1}. In contrast, relying on BinaryAI would have incurred significant verification overhead due to false positives. To investigate the details of vulnerability propagation, we subsequently compared the source code of decompress.c between \texttt{bzip2} 1.0.6 and 1.0.7. This analysis revealed that version 1.0.6---integrated within \texttt{lzbench 2.0.1}---omits a critical upper-bound check on the variable \textit{nSelectors}, as shown below:

\begin{lstlisting}[
    language=C,
    basicstyle=\ttfamily\small,
    % 关键设置：定义逃逸字符，这里使用 (*@ 和 @*) 作为界定符
    frame=single,
    breaklines=true,
    columns=flexible,
    escapeinside={(*@}{@*)} 
]
#define BZ_G_SIZE   50
#define BZ_MAX_SELECTORS (2 + (900000 / BZ_G_SIZE)) 
UChar selectorMtf[BZ_MAX_SELECTORS];

...

GET_BITS(BZ_X_SELECTOR_1, nGroups, 3);
if (nGroups < 2 || nGroups > 6) RETURN(BZ_DATA_ERROR);

GET_BITS(BZ_X_SELECTOR_2, nSelectors, 15);

(*@\textcolor{red}{if (nSelectors < 1) RETURN(BZ\_DATA\_ERROR); // bzip2 1.0.6}@*)
(*@\textcolor{blue}{if (nSelectors < 1) RETURN(BZ\_DATA\_ERROR); // lzbench 2.0.1}@*)

for (i = 0; i < nSelectors; i++) {
    ...
    s->selectorMtf[i] = j; // out-of-bounds write
}
\end{lstlisting}

\noindent The absence of this check permits out-of-bounds writes to the \textit{selectorMtf} array, constituting the root cause of CVE-2019-12900. Consequently, our analysis confirms that \texttt{lzbench}, by incorporating the vulnerable \texttt{bzip2} code, inherits this flaw and is susceptible to exploitation. Ultimately, this case study demonstrates that \toolname{} not only minimizes false positives by filtering out deceptive reuse but also enables the accurate identification of inherited vulnerabilities, thereby validating its efficacy in realistic scenarios.

\section{Related Work} \label{sec:related_word}

\textbf{Binary Code Similarity Calculation}:
Binary similarity calculation plays a crucial role in software reuse detection~\cite{Libdx, ISRD, Libdb, Libam}, malware analysis~\cite{BinSim, Malware 1, Malware 2, Malware 3}, vulnerability discovery~\cite{Bug Search 1, Bug Search 2, Bug Search 3, Bug Search 4, Bug Search 5, ACSAC 1, ACSAC 2, ACSAC 3, ACSAC 4, ACSAC 5}, and patch detection~\cite{Patch Detection 1, Patch Detection 2, Patch Detection 3, Patch Detection 4, Patch Detection 5, Patch Detection 6}, generally struggles to achieve an optimal balance between accuracy and computational efficiency.
Text hash–based approaches such as Gitz~\cite{Hash Based 1} and VIVA~\cite{Hash Based 2} are computationally efficient but fragile in the presence of complex binary compilation settings, leading to accuracy loss.
In contrast, symbolic execution techniques such as BinSim~\cite{BinSim} and Bingo~\cite{Bingo} enhance precision but are prone to path explosion, thereby incurring prohibitive computational overhead.
Recently proposed deep learning methods~\cite{ML Based 1, ML Based 2, ML Based 3} demonstrate strong representational capacity but demand extensive training data and substantial computing resources, constraining their practical efficiency. 
Specifically, code embedding-based solutions~\cite{NLP 1, NLP 2} learn token- and context-level code representations for tasks such as code completion and defect prediction.
In contrast, graph-based techniques using GNNs~\cite{GNN 1, GNN 2} capture structural semantics by modeling programs as abstract syntax trees or control/data-flow graphs.
Despite distinct modeling paradigms, both approaches suffer from notable efficiency issues, including high computational and memory overhead, slow training and inference, and limited scalability to large real-world codebases or incremental updates.

In this work, \toolname{} employs a lightweight Markov matrix encoding scheme in conjunction with a graph convolutional network to identify homologous functions, which significantly improves computational efficiency.
In addition, \toolname{} incorporates an novel area alignment strategy based on AFCGs to accurately locate homologous areas, thereby providing robust and reliable support for its subsequent reuse discrimination module.
Ultimately, these synergistic designs allow \toolname{} to balance efficiency and performance, ensuring more reliable TPL detection.

\textbf{TPL Detection}:
TPLs are widely used in modern software development, but insecure TPLs can introduce vulnerabilities, making accurate reuse detection essential. 
Existing TPL detection approaches can be broadly categorized into binary-to-source methods, which attempt to match binaries with their source code, and binary-to-binary methods, which directly compare binaries without relying on source availability.
Research on binary-to-source TPL detection, such as BinaryAI~\cite{BinaryAI} and TPLite~\cite{TPLite}, employs transformer-based matching and graph-assisted function-level origin detection, respectively. However, their reliance on binary-to-source mapping makes them impractical when source code is unavailable—a common scenario in software binary with proprietary and closed-source components—thus limiting their applicability in real-world TPL detection.
Recent methods for binary-to-binary detection such as BAT~\cite{BAT}, OSSPolice~\cite{OSSPolice}, and B2SFinder~\cite{B2sfinder} use constant-based fingerprints, which are fragile when such constants are absent. Other techniques like ISRD~\cite{ISRD} and LibDB~\cite{Libdb} rely on function similarity but suffer under compilation variance. To improve precision, ISRD uses similarity thresholds and LibDB utilizes FCG constraints, yet both incur high false positive/negative rates due to reliance on isolated matches. LibAM~\cite{Libam} improves accuracy by identifying reuse areas within the FCG, though its rigid alignment reduces recall. 

However, existing TPL detection methods often overlook the directionality of code reuse between code areas, hindering the effective distinction between real and deceptive reuse and eading to false positives in vulnerability assessments. While LibDI~\cite{LibDI} has explored complex reuse relationships in TPL, its proposed solution, which relies on function names, strings, and exported functions, suffers from robustness issues. In this work, we systematically analyze real and deceptive reuse cases, examine the topological differences between their corresponding code areas, and exploit these differences for accurate reuse discrimination.

\section{Conclusion}

In this study, we present \toolname{}, a multi-stage framework that simultaneously supports reuse discrimination and TPL detection. Unlike existing methods that often suffer from high false positive rates, \toolname{} redefines code reuse analysis by focusing on topological relationships within binaries—specifically, the centrality and connectivity of functions with respect to exported interfaces.

\toolname{} operates in three stages: it first identifies homologous functions using a lightweight Markov matrix encoding combined with a CNN; it then constructs AFCGs and applies a GCN to align homologous areas; finally, it quantifies topological differences between aligned areas to distinguish real from deceptive reuse, thereby enhancing TPL detection performance by reducing false positives.

To validate its effectiveness, we curated a dataset comprising deceptive reuse cases and conducted comprehensive comparisons against state-of-the-art techniques including LibAM, LibDB, and ISRD. 
Experimental results show that \toolname{} achieves a reuse discrimination F1 score of 96.9\%, substantially reducing false positives and achieving a TPL detection F1 score of 97.7\%---outperforming the best prior method by 9.2\%.
Moreover, \toolname{} maintains strong generalization capabilities across different architectures and compiler optimizations, with average F1 scores of 92.9\% and 92.0\%, respectively. Furthermore, comparisons with the commercial code analysis tool BinaryAI demonstrate that \toolname{}, by mitigating deceptive reuse, enables superior TPL detection and thereby enhances vulnerability detection.

\section{Limitations and Future Work}

While \toolname{} demonstrates robustness against heavily stripped binaries by leveraging intrinsic topological invariants rather than volatile string constants or symbol tables, we acknowledge specific boundaries in our current implementation. First, the tool exhibits a minor rate of false positives arising from topological ambiguities in donor-recipient relationships, particularly involving minute code fragments or generic utility functions with low topological distinctiveness. A representative instance of this ambiguity occurs between the donor \texttt{zlib} and its recipients, \texttt{minizip} and \texttt{lzbench}. The extensive invocation of generic \texttt{zlib} utilities within \texttt{minizip} artificially inflates centrality scores and creates an export-affinity that rivals the original library. Consequently, the model may misidentify the recipient \texttt{minizip} as the primary donor for \texttt{lzbench}, confounding a deceptive reuse relationship with a real one. Although such high-affinity pairings constitute a marginal fraction of our results (less than 2\% in our evaluation), they represent a precision bottleneck in handling generic libraries. Second, \toolname{} relies on the structural stability of the binary. Consequently, its detection accuracy faces challenges when analyzing binaries subjected to aggressive transformations, such as Link-Time Optimization (LTO) or heavy obfuscation. Specifically, LTO introduces cross-module inlining that fundamentally alters function boundaries, while obfuscation techniques deliberately distort execution logic. These mutations disrupt the locality assumptions essential to \toolname{}, necessitating specialized handling beyond standard structural analysis.

Driven by these identified limitations, our future research will focus on two strategic directions to enhance the tool’s fidelity and applicability. To mitigate the topological ambiguities observed in generic utility functions, we aim to integrate Large Language Models (LLMs) into our detection pipeline. By leveraging the semantic understanding and cross-domain reasoning capabilities of LLMs, we plan to complement topological features with semantic analysis, thereby enabling \toolname{} to distinguish between real reuse and structural coincidences in complex, low-distinctiveness scenarios. Simultaneously, to address the challenges posed by LTO and obfuscation, we plan to develop specialized preprocessing layers focused on de-obfuscation and structural normalization. This involves exploring hybrid signatures that persist despite aggressive inlining. To support the rigorous validation of these improvements, we will also automate the currently manual verification process of TPL reuse relationships. This automation will facilitate the generation of large-scale datasets, allowing for comprehensive stress-testing of \toolname{} against increasingly complex and obfuscated targets.






\end{document}